\documentclass[conference]{IEEEtran}
\ifCLASSINFOpdf
    \usepackage[pdftex]{graphicx}
\else
\fi
\usepackage{threeparttablex}

\usepackage{amsmath}
\ifCLASSOPTIONcompsoc
  \usepackage[caption=false,font=normalsize,labelfont=sf,textfont=sf]{subfig}
\else
  \usepackage[caption=false,font=footnotesize]{subfig}
\begin{document}
\title{RISC-V Based TinyML Accelerator for \\ Depthwise Separable Convolutions in Edge AI}


\author{
\IEEEauthorblockN{Muhammed Yildirim}
\IEEEauthorblockA{
Computer Engineering Department\\
Ihsan Dogramaci Bilkent University\\
06800 Ankara, Turkey\\
Email: m.yildirim@bilkent.edu.tr}
\and
\IEEEauthorblockN{Ozcan Ozturk}
\IEEEauthorblockA{
Computer Science and Engineering\\
Sabancı University\\
34956 Istanbul, Turkey\\
Email: ozcan.ozturk@sabanciuniv.edu}
}


%Telephone: (800) 555--1212, Fax: (888) 555--1212}

\maketitle

\begin{abstract}
The increasing demand for on-device intelligence in Edge AI and TinyML applications requires the efficient execution of modern Convolutional Neural Networks (CNNs). While lightweight architectures like MobileNetV2 employ Depthwise Separable Convolutions (DSC) to reduce computational complexity, their multi-stage design introduces a critical performance bottleneck inherent to layer-by-layer execution: the high energy and latency cost of transferring intermediate feature maps to either large on-chip buffers or off-chip DRAM. To address this memory wall, this paper introduces a novel hardware accelerator architecture that utilizes a fused pixel-wise dataflow. Implemented as a Custom Function Unit (CFU) for a RISC-V processor, our architecture eliminates the need for intermediate buffers entirely, reducing the data movement up to 87\% compared to conventional layer-by-layer execution. It computes a single output pixel to completion across all DSC stages-expansion, depthwise convolution, and projection-by streaming data through a tightly-coupled pipeline without writing to memory. Evaluated on a Xilinx Artix-7 FPGA, our design achieves a speedup of up to 59.3x over the baseline software execution on the RISC-V core. Furthermore, ASIC synthesis projects a compact 0.284 mm$^2$ footprint with 910 mW power at 2 GHz in 28 nm, and a 1.20 mm$^2$ footprint with 233 mW power at 300 MHz in 40 nm. This work confirms the feasibility of a zero-buffer dataflow within a TinyML resource envelope, offering a novel and effective strategy for overcoming the memory wall in edge AI accelerators.
\end{abstract}


% \fi
\IEEEpeerreviewmaketitle

% \section{Introduction}
%This demo file is intended to serve as a ``starter file''
%for IEEE conference papers produced under \LaTeX\ using
%IEEEtran.cls version 1.8b and later.
%I wish you the best of success.

 

%\subsection{Subsection Heading Here}
%Subsection text here.

%\subsubsection{Subsubsection Heading Here}
%Subsubsection text here.
%-----------------------------------------------------------
\section{Introduction}
%-----------------------------------------------------------
Edge AI\cite{singh2023edge} involves running artificial intelligence algorithms directly on local hardware, such as sensors and Internet of Things (IoT) units, bringing computation to the source of data creation. This allows for real-time processing without constant reliance on the cloud, an approach that offers several key benefits: low latency due to local processing, enhanced privacy by keeping sensitive data on the device, and reduced network bandwidth consumption, which enables reliable offline operation.\cite{yang2017method} A critical subfield of this domain is Tiny Machine Learning (TinyML)\cite{tinyml}, which specifically focuses on deploying machine learning models directly onto low-cost, ultra-low-power microcontrollers (MCUs) and embedded systems. These devices operate under severe constraints, often with power budgets in the milliwatt range and with only a few hundred kilobytes of memory, making on-device intelligence a significant technical challenge. The typical TinyML workflow involves taking a fully trained model and optimizing it for on-device inference by applying techniques such as quantization and pruning to create a smaller, more efficient model in a compact format.
%-----------------------------------------------------------

At the heart of modern computer vision lie Convolutional Neural Networks (CNNs)\cite{simonyan2014very, szegedy2016rethinking,he2016deep}, but their computational and memory intensity presents a formidable barrier to their deployment on edge devices. To bridge this gap, researchers have developed architecturally efficient models like Efficientnet\cite{tan2019efficientnet}, Xception\cite{chollet2017xception}, ShufflenetV2\cite{ma2018shufflenet}, MobileNetV2\cite{MobileNetV2}, which utilizes Depthwise Separable Convolutions (DSC) to dramatically reduce the number of parameters and operations.
%-----------------------------------------------------------

Despite these algorithmic optimizations, a fundamental system-level challenge remains: the memory wall. While Depthwise Separable Convolutions (DSCs) reduce arithmetic operations, the structure of their core building block in models like MobileNetV2—the inverted residual block—creates a new bottleneck in a conventional layer-by-layer execution model. In this model, the entire output of one layer, an "intermediate feature map," must be written to memory before the next layer can begin its computation. This means that after an operation like a depthwise convolution is complete, its results must be transferred to either a large on-chip buffer or off-chip DRAM, only to be read back in for the next step.\cite{li2021dynamic}
%-----------------------------------------------------------

Moving this data constantly is very costly. For example, in our measurements of a representative layer in TensorFlow Lite (TFLite) implementation of MobileNetV2 with 8-bit data, just storing the intermediate feature map used about 38 KB of on-chip SRAM. Previous research shows that moving data uses much more energy than performing the actual calculations. Transferring data between on-chip and off-chip memory, and within the chip itself, costs far more than the multiply-accumulate (MAC) operations \cite{chen2016eyeriss}\cite{yang2017method}. In the TinyML context, where microcontrollers often have 256 KB of SRAM or less and no DRAM, every kilobyte of memory and milliwatt of power is critical. Consequently, moving this intermediate data back and forth becomes the main factor limiting both performance and battery life \cite{lin2023tiny}.
%-----------------------------------------------------------

Hardware acceleration is the natural approach to overcoming these performance limitations, but the existing landscape reveals a critical gap. Numerous accelerators have been proposed to optimize DSCs, often employing sophisticated dataflows and layer fusion techniques to reduce memory traffic. However, these designs are typically implemented on large, high-end FPGAs, consuming significant resources and power that make them unsuitable for the TinyML domain \cite{liu2019fpga} \cite{bai2018cnn}\cite{wu2019high} \cite{ding2019designing} \cite{fan2018real} \cite{li2021dynamic} \cite{ xuan2022fpga} \cite{li2022efficient}\cite{huang2023high} \cite{li2024high}. On the other hand, existing TinyML accelerators do operate within the correct small power and area budgets. However, they generally focus on accelerating individual, specific operations—like custom instructions for matrix math, special units for sparse calculations, or accelerators for single convolution types. These accelerators typically run in a conventional layer-by-layer model, where software (like the TFLite interpreter) tells the hardware which operation to run next. Even TinyML designs that do focus on the memory problem aim to make the on-chip buffers smaller or smarter, not to eliminate them.\cite{prakash2023cfu} \cite{wu2024ai}\cite{balasubramaniam2025effective}\cite{lin2022low}\cite{gao2023liteair5}\cite{haris2023secda}\cite{ sabih2024hardware}\cite{manor2022custom}. This leaves the system-level bottleneck of intermediate feature map buffering unaddressed.
%-----------------------------------------------------------

This paper directly attacks this memory wall. We introduce a new hardware accelerator that uses a fused pixel-wise dataflow to eliminate the memory bottleneck between layers in DSC blocks. Implemented as a Custom Function Unit (CFU) for a RISC-V processor, our architecture re-organizes the computation from a layer-by-layer model to a pixel-by-pixel model. It computes a single output pixel completely—across all DSC stages of expansion, depthwise convolution, and projection—by streaming data through a tightly-coupled pipeline without writing intermediate feature maps to memory.

This approach is designed specifically for the TinyML domain, focusing on memory and energy savings to allow complex models to run on resource-constrained hardware. The main contributions of this work are:
\begin{itemize}
    \item A fused pixel-wise dataflow architecture that completely removes the need for on-chip SRAM buffers for intermediate feature maps, reducing a major source of area, latency, and power use.

    \item A co-designed hardware architecture that includes a three-stage parallel pipeline and a specialized memory system with on-the-fly padding logic, all carefully designed to enable and support the fine-grained, zero-buffer dataflow.

    \item A full-stack implementation and validation of the accelerator as a RISC-V CFU on a low-cost FPGA, showing that this advanced dataflow strategy is possible within the tight resource limits of TinyML. Our complete evaluation, including both FPGA measurements and 40nm/28nm ASIC synthesis, confirms large speedup and energy efficiency gains over a basic software-only implementation.
\end{itemize}
%-----------------------------------------------------------
\section{Background}
%-----------------------------------------------------------
\subsection{Depthwise Separable Convolutions (DSC)}
\begin{figure}[htbp]
    \centering
    \subfloat[Standard Convolution]{
        \includegraphics[width=0.9\columnwidth]{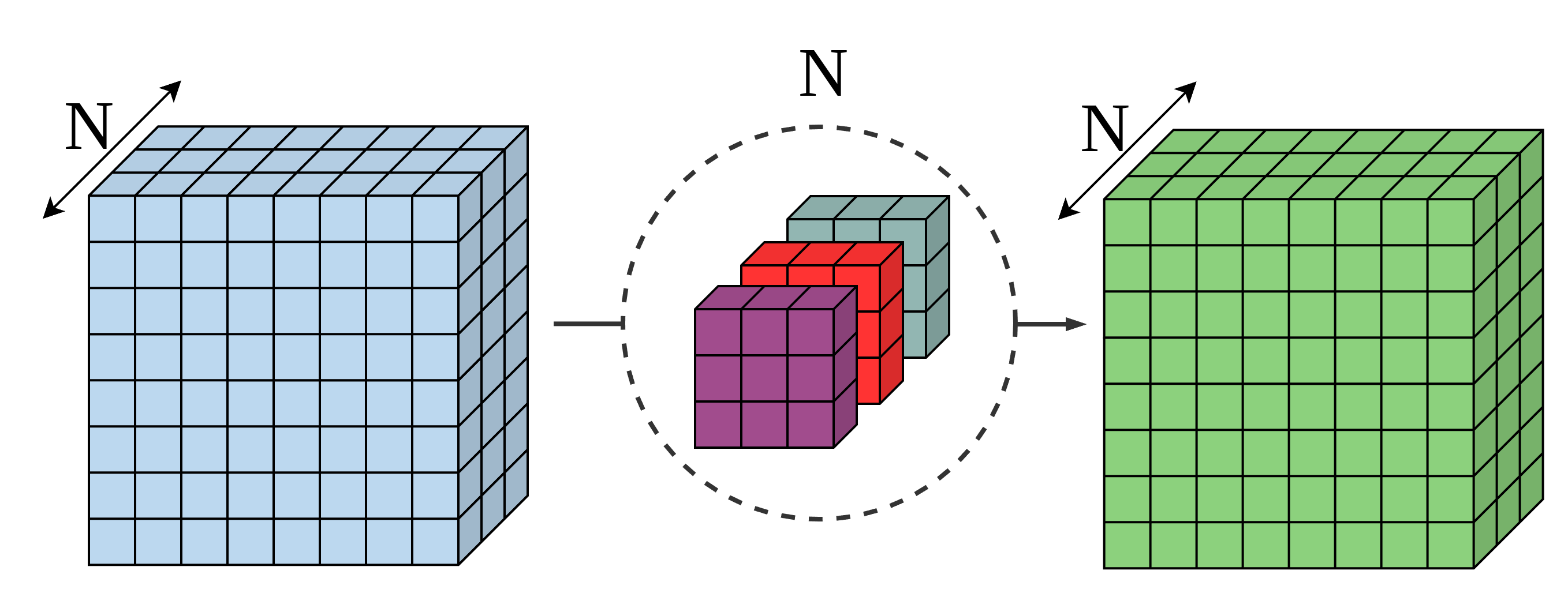}
        \label{fig:standart_conv}
    }
    \\ % This creates the vertical break
    \subfloat[Depthwise Separable Convolution]{
        \includegraphics[width=0.8\columnwidth]{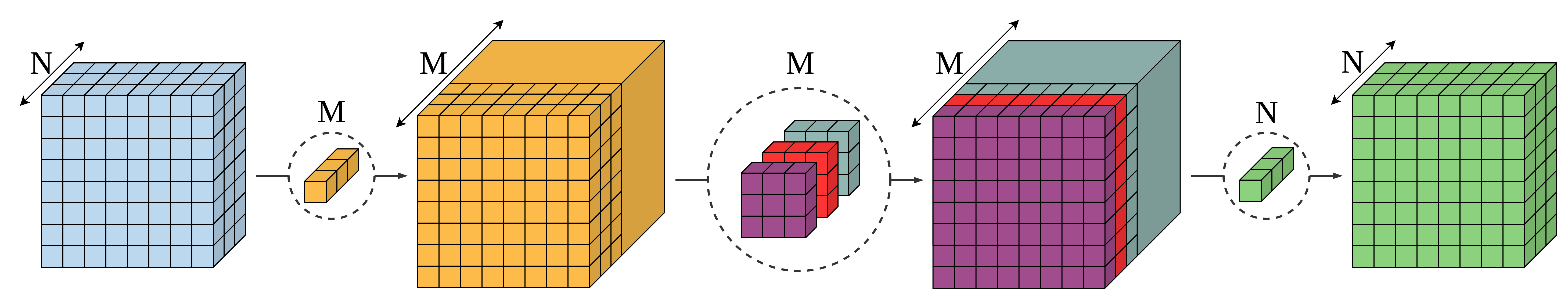}
        \label{fig:depthwise_separable_conv}
    }
    \caption{MobileNetV2 - residual block: Comparison of standard convolution (top) and depthwise separable convolution (bottom). DSC factorizes the operation into depthwise and pointwise stages, significantly reducing computation and parameter count.}
    \label{fig:convolutions}
\end{figure}

%-----------------------------------------------------------
Depthwise Separable Convolution (DSC) is a factorized form of convolution that serves as the cornerstone of many lightweight Convolutional Neural Networks (CNNs), including \textit{MobileNetV1} and \textit{MobileNetV2}. As illustrated in Figure~\ref{fig:convolutions}, it was introduced to dramatically reduce computational cost and model size by decoupling spatial and channel-wise correlations into two simpler steps: a depthwise convolution (DWC) and a pointwise convolution (PWC).

MobileNetV2, which is the target model of this work, integrates DSC into a specific three-stage structure called the \textbf{inverted residual block}, consisting of:

\begin{enumerate}
    \item An initial 1x1 "Expansion" convolution that increases the number of channels.
    \item A 3x3 Depthwise convolution that performs spatial filtering on the expanded feature map.
    \item A final 1x1 "Projection" convolution that reduces the channels back down, forming a bottleneck.
\end{enumerate}

The efficiency of this factorization is evident when comparing the number of Multiply-Accumulate (MAC) operations against a standard convolution. For an input feature map of size $W \times W \times M$ and an output of $W \times W \times N$ with a $K \times K$ kernel, the costs are:
\begin{itemize}
    \item \textbf{Standard Convolution (SC):} \\
    $O_{SC} = W \times W \times K \times K \times M \times N$
    \item \textbf{Depthwise Separable Convolution (DSC):} \\
    $O_{DSC} = (W \times W \times K \times K \times M) + (W \times W \times M \times N)$
\end{itemize}

This factorization yields a significant reduction in both operations and parameters, which can be expressed as a ratio:
$$\frac{O_{DSC}}{O_{SC}} = \frac{1}{N} + \frac{1}{K^2}$$
For a typical $3\times3$ kernel ($K=3$), DSC requires roughly 8–9$\times$ fewer operations than a standard convolution, making it highly suitable for \textit{resource-constrained edge and TinyML devices}.
%-----------------------------------------------------------
%-----------------------------------------------------------
\subsection{Custom Functional Units \& CFU-Playground}
\begin{figure}[htbp]
    \centering
    \includegraphics[width=0.9\columnwidth]{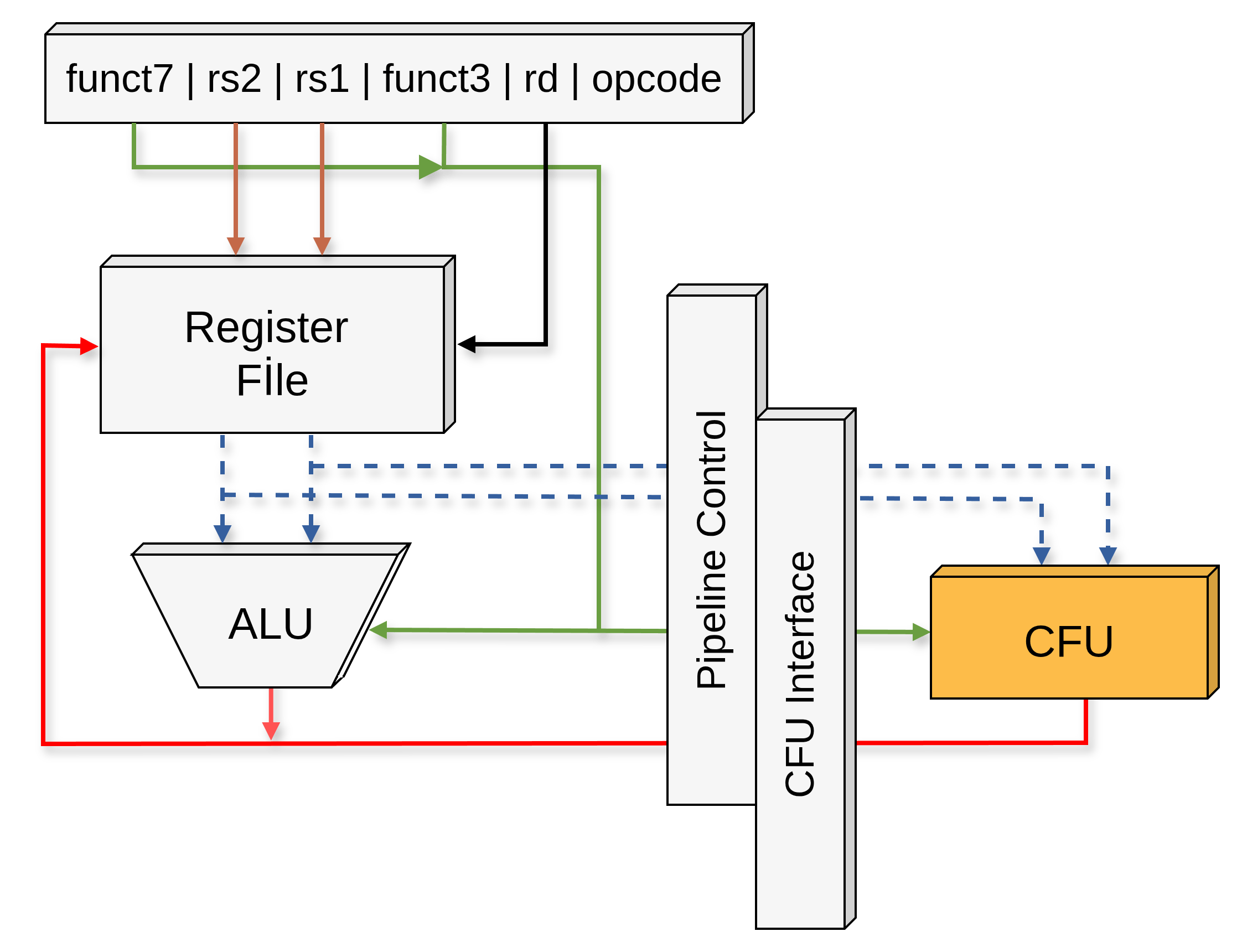} 
    \caption{CPU-CFU interface using R-type instruction of RISC-V ISA in CFU-Playground framework.}
    \label{fig:cfu_playground_figure}
\end{figure}

%-----------------------------------------------------------
Custom Function Units (CFUs) and the CFU-Playground framework provides a flexible environment for developing and testing custom hardware accelerators integrated with RISC-V processors. CFUs are specialized hardware modules that extend the capabilities of a RISC-V core by implementing custom instructions tailored for specific computational tasks. This allows for offloading intensive operations from the main processor, thereby improving performance and energy efficiency. The CFU-Playground framework simplifies the process of designing, simulating, and deploying these custom units. It provides a modular architecture where developers can define their CFU's functionality, interface it with the RISC-V core, and test it using a variety of tools. Figure~\ref{fig:cfu_playground_figure} illustrates the integration of a CFU with a RISC-V CPU using R-type instructions from the RISC-V ISA. This setup allows the CPU to invoke the CFU for specific operations, enabling efficient execution of tasks that benefit from hardware acceleration.
%-----------------------------------------------------------
%-----------------------------------------------------------
\subsection{DSC Accelerator Architectures}
\begin{figure}[htbp]
    \centering
    \includegraphics[width=0.9\columnwidth]{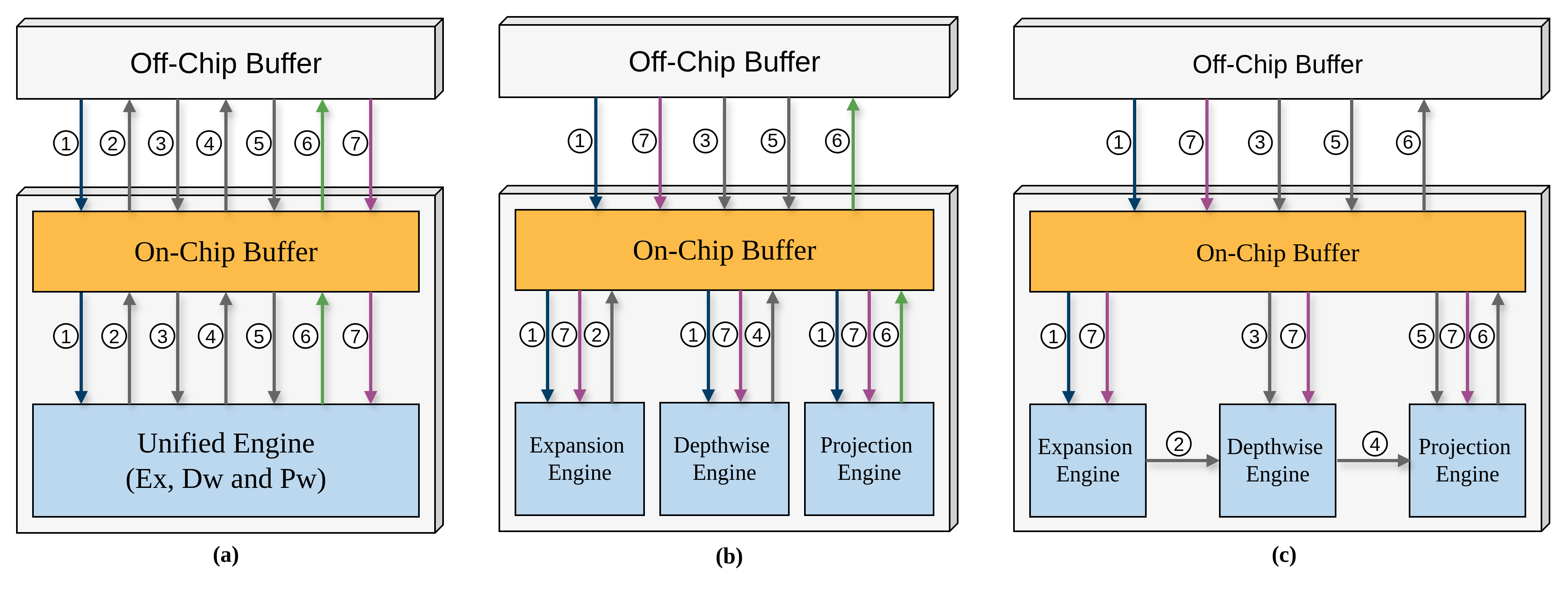} 
    \caption{Comparison of DSC accelerator architectures. (a) The Unified Architecture, which uses a single engine for all stages, creating a significant off-chip memory traffic bottleneck. (b) The Separated Architecture, which reduces off-chip traffic but introduces a new bottleneck with the large on-chip buffer required for inter-engine communication. (c) Our Proposed Fused Architecture, which eliminates the intermediate buffer entirely by streaming data directly between engines, solving both prior bottlenecks.}
    \label{fig:unified_separated}
\end{figure}

%-----------------------------------------------------------
While Depthwise Separable Convolutions (DSCs) greatly reduce arithmetic operations, their multi-stage structure (e.g., Expansion, Depthwise, and Projection in MobileNetV2) introduces a major challenge for hardware acceleration: the \textbf{memory wall}. In a conventional layer-by-layer execution, each stage must store its entire output feature map before the next stage can begin, resulting in large memory traffic and energy overhead. To mitigate this issue, accelerator architectures have evolved through several design patterns, as illustrated in Figure~\ref{fig:unified_separated}.

The most basic approach is the \textbf{Unified Architecture} (Figure~\ref{fig:unified_separated}(a)), which uses a single, general-purpose compute engine for all operations, often chosen for flexibility and high resource reuse \cite{liu2019fpga}. While adaptable (capable of processing both standard and depthwise convolutions within the same hardware), this design is severely bottlenecked by off-chip memory traffic, as it must write and read back every intermediate feature map from DRAM. Even designs using configurable engines \cite{bai2018cnn} within this unified paradigm still face significant challenges managing memory access and bandwidth limitations inherent to this layer-by-layer approach.

To reduce this bottleneck, Separated Architectures (Figure~\ref{fig:unified_separated}(b)) employ dedicated engines for different convolution types (e.g., one for depthwise, one for pointwise)~\cite{wu2019high}.This allows inter-layer pipelining~\cite{ding2019designing} and significantly reduces off-chip traffic by holding intermediate data in large on-chip SRAM buffers using techniques such as double-buffering~\cite{ding2019designing} or intra-block caching~\cite{fan2018real}. However, this introduces a new bottleneck: these on-chip buffers can be very large, consuming significant chip area and power~\cite{fan2018real}. Designs that use dynamic scheduling or runtime switching between DW and PW modes can improve engine utilization but still incur high memory traffic costs when intermediate results are frequently moved~\cite{li2021dynamic}. Recent hybrid dataflow architectures, such as Zhao~et~al.~\cite{zhao2025high}, employ multiple compute engines (CEs) and dynamic scheduling to balance utilization across DSC layers. Although this improves performance, they still require sizable intermediate buffers to synchronize inter-engine communication, which limits efficiency on resource-constrained devices.

The current state-of-the-art for high-performance systems is to move toward Fused Architectures (Figure~\ref{fig:unified_separated}(c)), which aim to further reduce or eliminate this intermediate buffer by improving dataflow between stages. Many accelerators have explored sophisticated dataflows to achieve partial or full fusion. For instance, Xuan et al. \cite{xuan2022fpga} developed a design that fuses the Depthwise and Projection stages, successfully reducing BRAM traffic by over 80\%. Guoqing Li et al. \cite{li2022efficient} proposed efficient shared PEs and dataflow strategies like Pointwise-to-Depthwise (P2D) and Depthwise-to-Pointwise (D2P) to minimize external memory access by keeping intermediate data on-chip between specific stages. Huang et al. \cite{huang2023high} also achieve partial fusion and propose efficient convolution algorithms specifically to reduce on-chip memory occupancy and improve off-chip access patterns. Similarly, Li et al.\cite{li202150} introduced a double-layer multiply-accumulate (MAC) pipeline that jointly processes the depthwise and projection layers in a streaming fashion, reducing BRAM traffic by 41\%. This approach demonstrates the benefits of partial in-block fusion and dynamic DSP allocation for improving energy efficiency. Other works explore pixel-level pipelines \cite{li2024high} to reduce memory access.

While these advanced architectures validate the system-level benefits of fusion, they are fundamentally designed for high-performance, high-resource FPGAs (e.g., Virtex-7, Zynq UltraScale+, Arria 10) and are unsuitable for the TinyML domain \cite{xuan2022fpga, li2021dynamic, li2024high, ding2019designing, huang2023high, li2022efficient}. Their architectures often consume hundreds of thousands of LUTs and significant power (e.g., 6-40W) \cite{xuan2022fpga, li2021dynamic, ding2019designing}, orders of magnitude too large for the milliwatt and kilobyte-scale constraints of an edge device. Crucially, while they significantly reduce intermediate buffering through techniques like partial fusion or P2D/D2P, they often do not completely eliminate it for the full Ex $\rightarrow$ Dw $\rightarrow$ Pr sequence, still relying on some on-chip SRAM to store intermediate rows, tiles, or feature maps between stages \cite{li2021dynamic, fan2018real, huang2023high}. This highlights a critical gap: the idea of layer-fusion is proven effective for performance, but a TinyML-compatible, zero-buffer implementation is missing.

%-----------------------------------------------------------
%-----------------------------------------------------------
\subsection{TinyML Accelerators}
%-----------------------------------------------------------
While high-performance accelerators tackle the memory wall through complex dataflows on resource-rich platforms, the field of TinyML operates under drastically different constraints: ultra-low power (milliwatts), small memory footprints (kilobytes), and the need for flexible, low-cost deployment, often on MCU-class devices \cite{warden2019tinyml, lin2023tiny}. A prominent approach in this domain is hardware/software co-design, particularly leveraging the extensibility of the open-source RISC-V instruction set architecture (ISA) \cite{gao2023liteair5, sainz2024hardware}.

Frameworks like CFU Playground \cite{prakash2023cfu} facilitate this tightly-coupled co-design by enabling the creation of Custom Function Units (CFUs)–hardware accelerators invoked via custom RISC-V instructions. This allows developers to offload critical computational kernels while retaining software flexibility on the main CPU core. Alongside this CFU approach, numerous works have demonstrated the effectiveness of loosely-coupled accelerators communicating with the TensorFlow Lite Micro (TFLM) ecosystem, often as TFLM delegates connected via a system bus like AXI \cite{manor2022custom, haris2023secda}.

Many TinyML acceleration efforts focus on techniques that are different from dataflow fusion. For example, many designs focus on handling sparsity and reconfigurability\cite{krishna2024raman}, accelerating specific, individual operations like standard convolutions \cite{prakash2023cfu}, transposed convolutions \cite{wu2024ai}, or even use CFUs primarily as smart memory controllers \cite{balasubramaniam2025effective}, again underscoring the dominance of memory access costs in TinyML \cite{balasubramaniam2025effective}. Even accelerators designed for MobileNetV2 on resource-constrained FPGAs often adopt conventional layer-by-layer processing, relying on ping-pong buffers or other on-chip memory schemes to hold intermediate feature maps between DSC stages \cite{lin2022low}. While these designs fit within TinyML resource budgets, they do not directly address the intra-block memory bottleneck inherent to the multi-stage DSC structure itself.

This reveals a specific gap within TinyML acceleration: while CPU+CFU co-design is established, and techniques like sparsity and quantization are well-explored, the system-level memory bottleneck caused by intermediate feature maps within DSC blocks has largely been unaddressed by existing CFU-based approaches. Prior work has not demonstrated a zero-buffer, fused pixel-wise dataflow spanning the full Expansion→Depthwise→Projection sequence, implemented as a tightly-coupled RISC-V CFU specifically designed to eliminate this intermediate buffering within TinyML constraints. Our work aims to fill this gap, adapting the principle of layer fusion, previously confined to high-performance accelerators, into the resource-constrained, co-design paradigm of TinyML.
%-----------------------------------------------------------
\section{Proposed Architecture}
%-----------------------------------------------------------
\subsection{Fused Pixel-wise Dataflow}
\begin{figure*}[!t]
    \centering
    % --- Top Figure ---
    \subfloat[]{\includegraphics[width=0.85\textwidth]{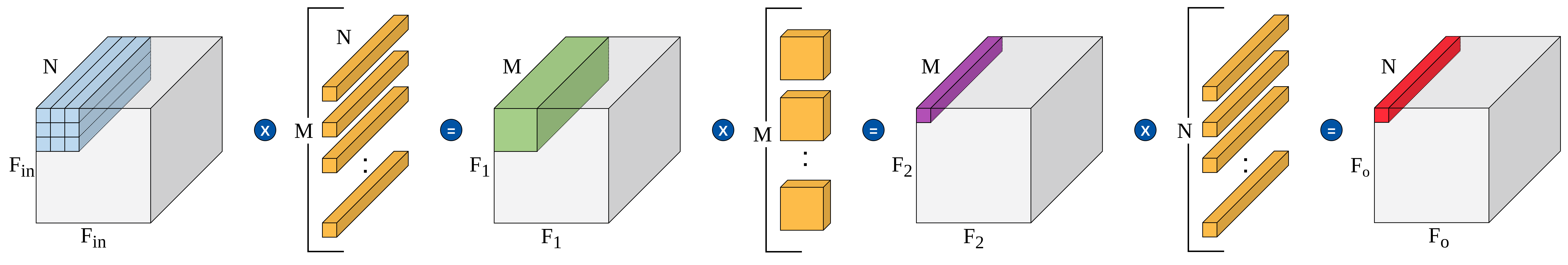}\label{fig:full_dataflow}}
    
    % --- Bottom Figure ---
    \subfloat[]{\includegraphics[width=0.85\textwidth]{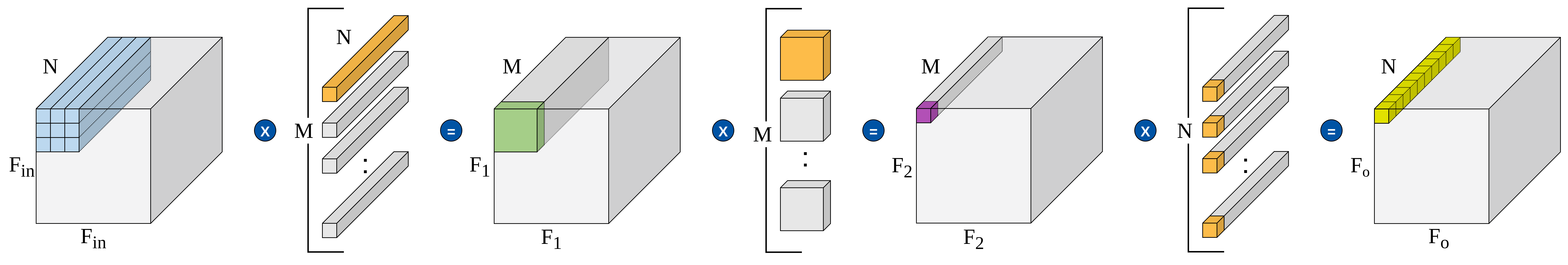}\label{fig:partial_dataflow}}
    
    \caption{The proposed fused pixel-wise dataflow. (a) Illustrates the complete computation for a single output pixel, showing the full interaction between input and output channels across the three convolution stages. (b) A granular view of the element-wise computation, highlighting how intermediate results are immediately streamed between stages to eliminate the need for intermediate buffering of the entire feature map.}
    \label{fig:dataflows}
\end{figure*}

%-----------------------------------------------------------
The inverted residual linear bottleneck blocks of the MobileNetV2~\cite{MobileNetV2} consist of two PWC layers and one DWC layer.  In the inference phase, one layer is calculated and then used as the input of the next layer. This sequential approach, however, requires either substantial memory traffic for the intermediate feature maps ($F_1$ and $F_2$) or a large on-chip memory footprint to store them. To overcome this limitation, our proposed architecture, illustrated in Figure~\ref{fig:dataflows} employs fused pixel-wise dataflow, that eliminates both intermediate buffering and off-chip memory access.

The core of this dataflow is the re-organization of the computation from a layer-centric to a pixel-centric approach. Instead of computing the entire feature map for one layer before proceeding to the next , our pipeline is structured to produce one complete output pixel before moving on to the next. As shown in Figure~\ref{fig:partial_dataflow} this is achieved by tightly coupling the computational stages of Expansion, Depthwise, and Projection. The pipeline is designed to produce only the minimum amount of data necessary for the subsequent stage to begin its work; generating more would be unnecessary and require the large on-chip buffers we seek to avoid. For example, once the Expansion stage has produced a 3x3x1 tile of the intermediate feature map ($F_1$), this is sufficient for the Depthwise stage to compute a single 1x1x1 element of $F_2$. This single element is, in turn, immediately streamed to the Projection stage. This deep, synchronous pipeline ensures that the intermediate results, $F_1$, and $F_2$ , exist only as transient data within the hardware registers for a few clock cycles.

The benefits of the fused dataflow can be evaluated by comparing its memory requirements with those of conventional non-fused approaches. In a traditional layer-by-layer system, intermediate results must be stored in off-chip DRAM, leading to substantial memory traffic is
\begin{equation}
    \text{Traffic}_{\text{DRAM}} = 2 \times (H_1 \times W_1 \times C_1) + 2 \times (H_2 \times W_2 \times C_2)
\end{equation}
Even a more optimized pipelined design that avoids DRAM still requires a large on-chip buffer to store intermediate feature maps, with a minimum size of 
\begin{equation}
    \text{Buffer}_{\text{SRAM}}  = (H_1 \times W_1 \times C_1)
\end{equation}
For the fifth bottleneck layer of MobileNetV2 (TensorFlow Lite), where both intermediate feature maps ($F_1$ and $F_2$) are sized $20\times20\times96$, this would necessitate over 153 KB of off-chip traffic or a 38.4 KB on-chip buffer. In contrast, our fused pixel-wise dataflow eliminates both the intermediate off-chip traffic and the on-chip buffer requirement to zero, directly translating to significant savings in latency, power, and area.
%-----------------------------------------------------------
%-----------------------------------------------------------
\subsection{Dedicated Engine Architectures}
\begin{figure}[htbp]
	\centering
	\includegraphics[width=0.4\textwidth]{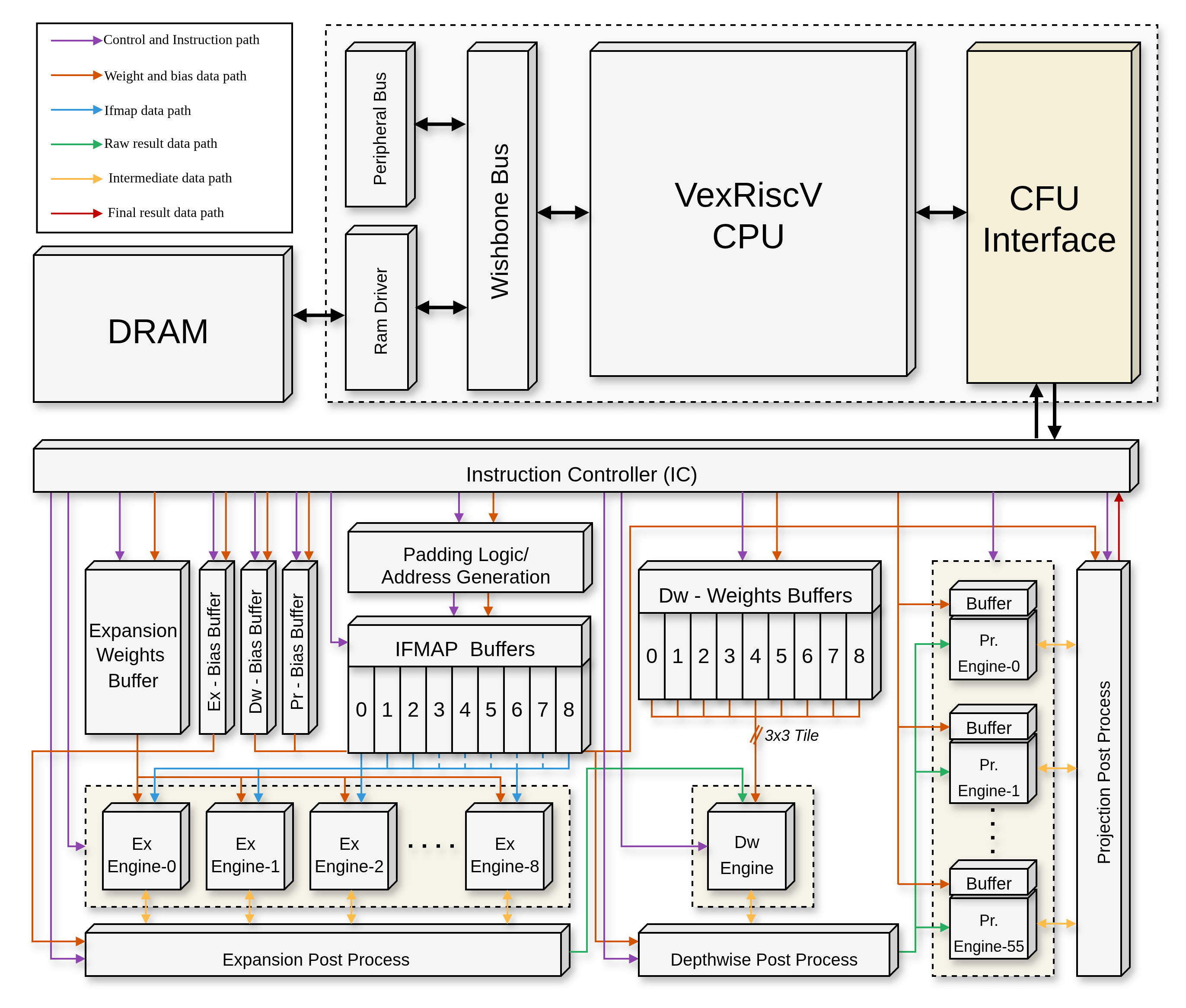}
	\caption{High-level block diagram of the proposed fused DSC accelerator. The architecture features a three-stage pipeline with dedicated parallel engines for the Expansion, Depthwise, and Projection stages, orchestrated by an Instruction Controller. Specialized memory structures, such as the parallel IFMAP and Dw-Weights Buffers, are designed to maximize data throughput. The legend details the color-coded paths for control, weights, IFMAP, and intermediate data flowing through the system.}
	\label{fig:full_design}
\end{figure}

%-----------------------------------------------------------
Our accelerator is implemented as a \textbf{Custom Function Unit (CFU)} tightly coupled with a \textbf{VexRiscv}\cite{spinalhdl_vexriscv_2023} processor within the LiteX SoC framework, as shown in Figure~\ref{fig:full_design}. The CFU core consists of a three-stage pipeline, with each stage implemented as a dedicated compute engine: the Expansion Unit, the Depthwise Unit, and the Projection Unit. This organization allows each engine to be optimized for its specific operation (e.g., $1{\times}1$ or $3{\times}3$ convolutions) and corresponding data-access patterns. An Instruction Controller (IC) orchestrates the dataflow and synchronization between stages, ensuring that results are streamed directly from one engine to the next without intermediate buffering. To sustain this high-throughput pipeline, specialized on-chip memory subsystems are used, including parallel buffers for input feature maps (IFMAPs) and weights.

%-----------------------------------------------------------
%-----------------------------------------------------------

\begin{figure}[htbp]
    \centering
    \subfloat[Parallel Multiply-Accumulate Engines]{
        \includegraphics[width=0.9\columnwidth]{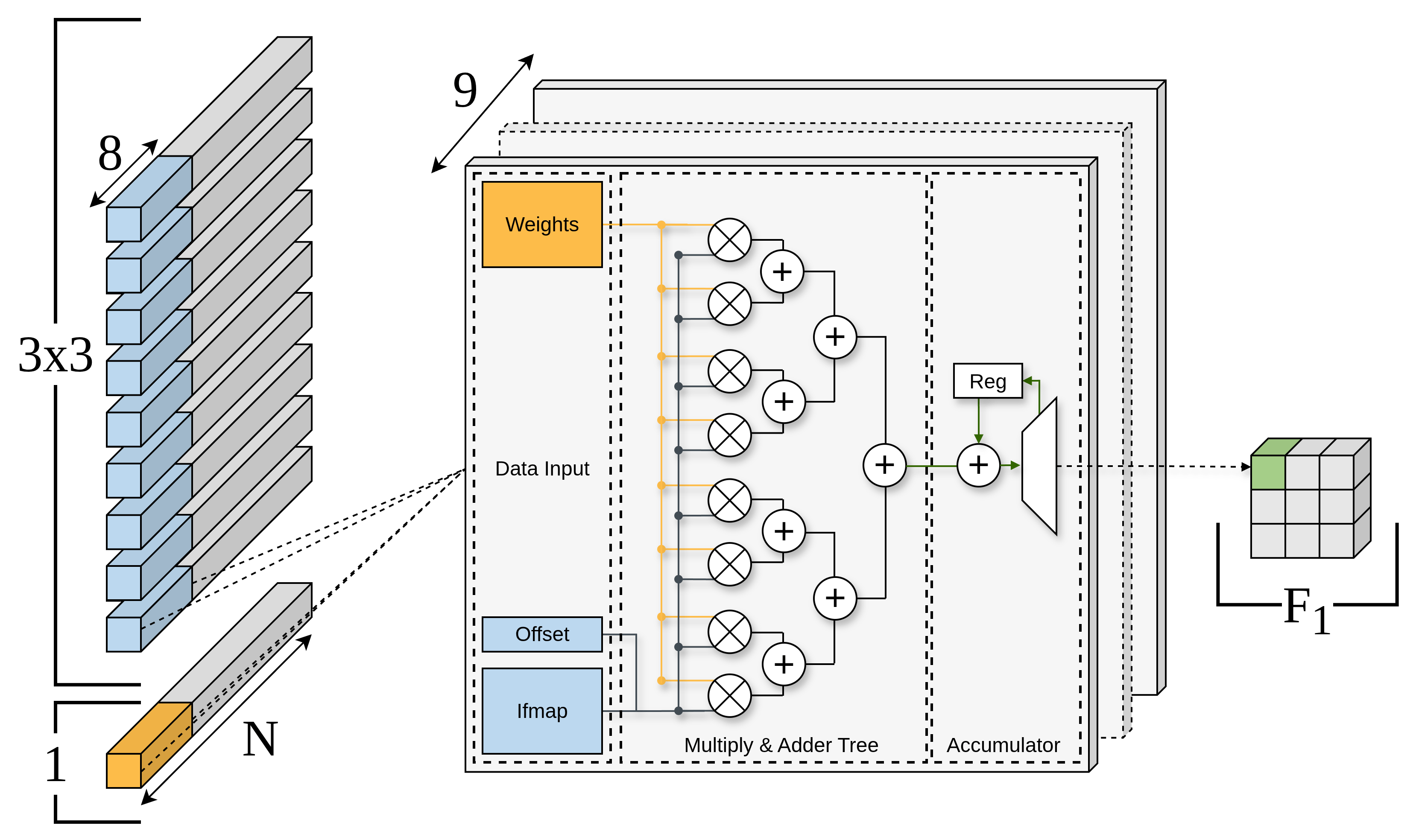}
        \label{fig:expansion_engine}
    }
    \\ % This creates the vertical break
    \subfloat[Post-Processing Pipeline]{
        \includegraphics[width=0.8\columnwidth]{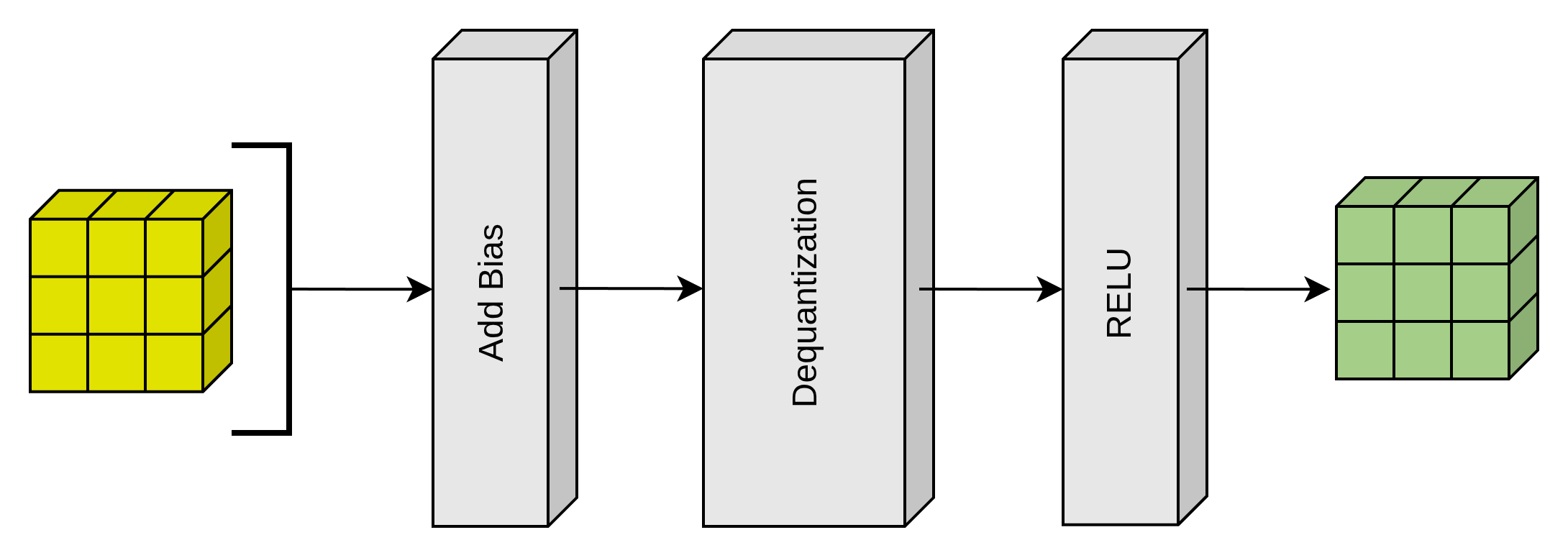}
        \label{fig:expansion_post_process}
    }
    \caption{Detailed architecture of the Expansion stage. (a) A $3 \times 3$ tile of the N-channel input feature map is processed by nine parallel Expansion Engines. Each engine computes a dot product between the input pixel's N channels and a filter's weights to produce a single raw output channel. (b) The raw results from the engines enter a post-processing pipeline for bias addition, requantization, and ReLU activation, producing one tile of the intermediate feature map ($F_1$).}
    \label{fig:expansion_stage_detail}
\end{figure}

The Expansion Unit is the first and most computationally intensive stage of the DSC pipeline, performing the initial $1{\times}1$ pointwise convolution that expands the feature map’s channel depth. As illustrated in Figure~\ref{fig:expansion_stage_detail}(a), this stage employs an Input-Stationary (IS) dataflow optimized for high spatial and data-level parallelism. A $3{\times}3$ tile of input data is fetched from the on-chip IFMAP buffer and held stationary across nine parallel Expansion Engines. The same expansion filter is broadcast to all engines so that, in each cycle, every engine computes one output pixel corresponding to one position in the $3{\times}3$ tile. Each engine performs a dot product between the $N$ input channels of its assigned pixel and the $N$ corresponding weights of the active filter, producing one channel of the intermediate feature map $F_1$. The computation proceeds channel by channel, keeping the IFMAP stationary while streaming different expansion filters through the engines. Inside each Expansion Engine, an eight-way multiply–accumulate (MAC) tree performs parallel operations on eight input channels per cycle. This organization fully exploits both spatial parallelism—across the nine engines—and data-level parallelism—within each engine. The choice of eight-way MAC parallelism ensures 100\% hardware utilization since all channel dimensions in MobileNetV2 are multiples of eight. At the end of the accumulation stage, the 32-bit results are forwarded to a post-processing pipeline, shown in Figure~\ref{fig:expansion_stage_detail}(b), where bias addition, dequantization, and ReLU activation are applied to produce the final 8-bit quantized outputs for the subsequent Depthwise Unit.
%-----------------------------------------------------------

%-----------------------------------------------------------

\begin{figure}[htbp]
	\centering
	\includegraphics[width=0.4\textwidth]{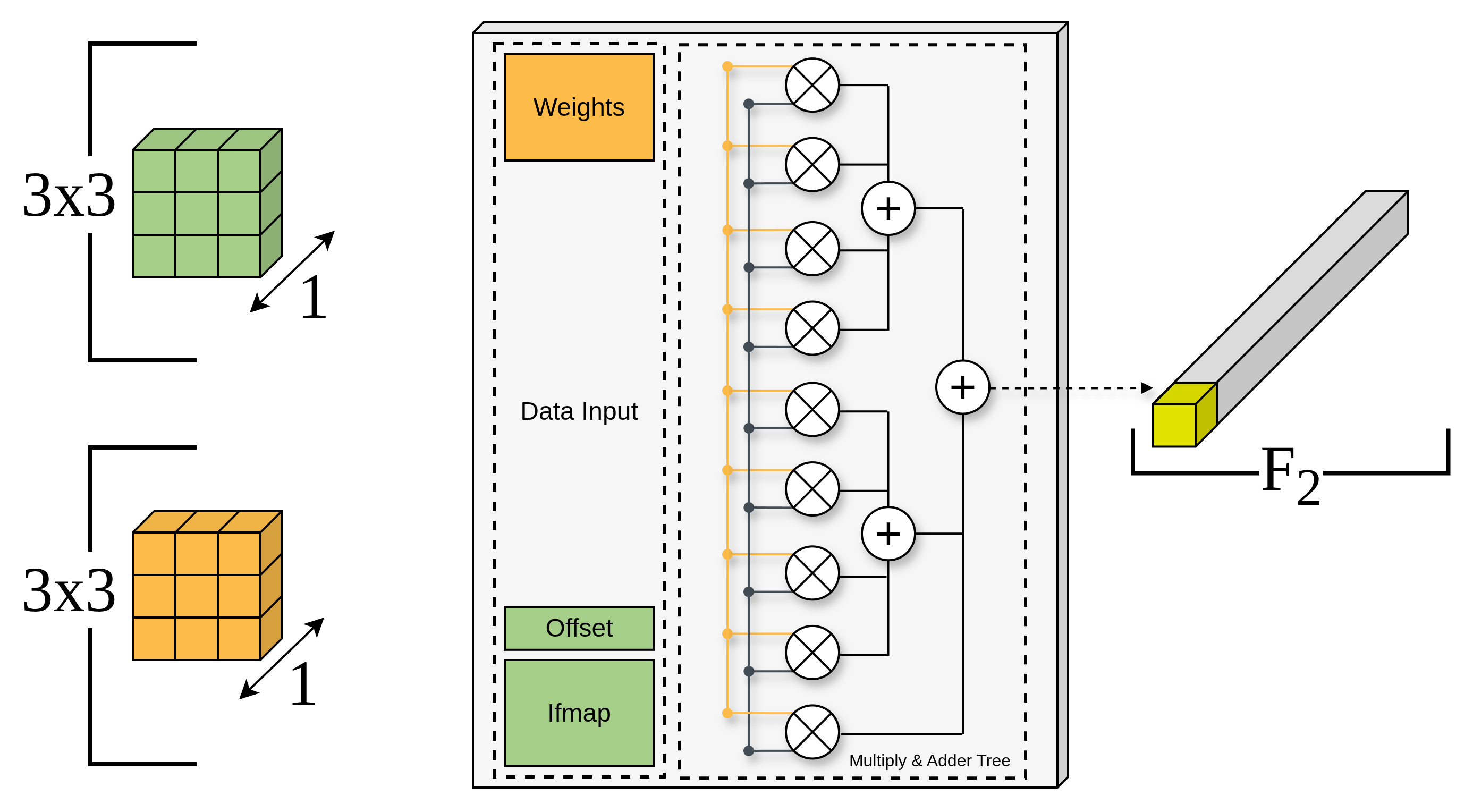}
	\caption{Internal architecture of the Depthwise Engine. A $3 \times 3$ spatial window from the intermediate feature map ($F_1$) is convolved with its corresponding $3 \times 3$ depthwise filter (Weights). The Multiply \& Adder Tree computes a single element for one channel of the second intermediate feature map ($F_2$).}
	\label{fig:depthwise_engine}
\end{figure}

The Depthwise Unit performs the $3{\times}3$ depthwise convolution on the $M$ channels of the intermediate feature map ($F_1$), serving as a streamlined pipeline stage between the Expansion and Projection units. As shown in Figure~\ref{fig:depthwise_engine}, this unit consists of a single, highly parallel Depthwise Engine optimized for minimal latency and continuous data streaming. It operates with a No Local Reuse (NLR) dataflow, meaning that for each channel, the engine fetches one $3{\times}3$ spatial window from $F_1$ along with its corresponding $3{\times}3$ depthwise filter, performs one convolution to produce a single output element of the next intermediate feature map ($F_2$), and then discards the inputs. This simple streaming approach prioritizes direct data movement over local buffering, enabling smooth synchronization with the preceding and succeeding stages. The Depthwise Engine integrates a nine-way multiply–accumulate (MAC) array, allowing all nine multiplications of a $3{\times}3$ convolution to occur within a single cycle, followed by an adder tree for accumulation. After computation, the raw 32-bit results are passed through a dedicated post-processing pipeline that applies bias addition, requantization, and ReLU activation before forwarding the processed data to the Projection Unit.
%-----------------------------------------------------------

%-----------------------------------------------------------

\begin{figure}[htbp]
    \centering
    \includegraphics[width=0.9\columnwidth]{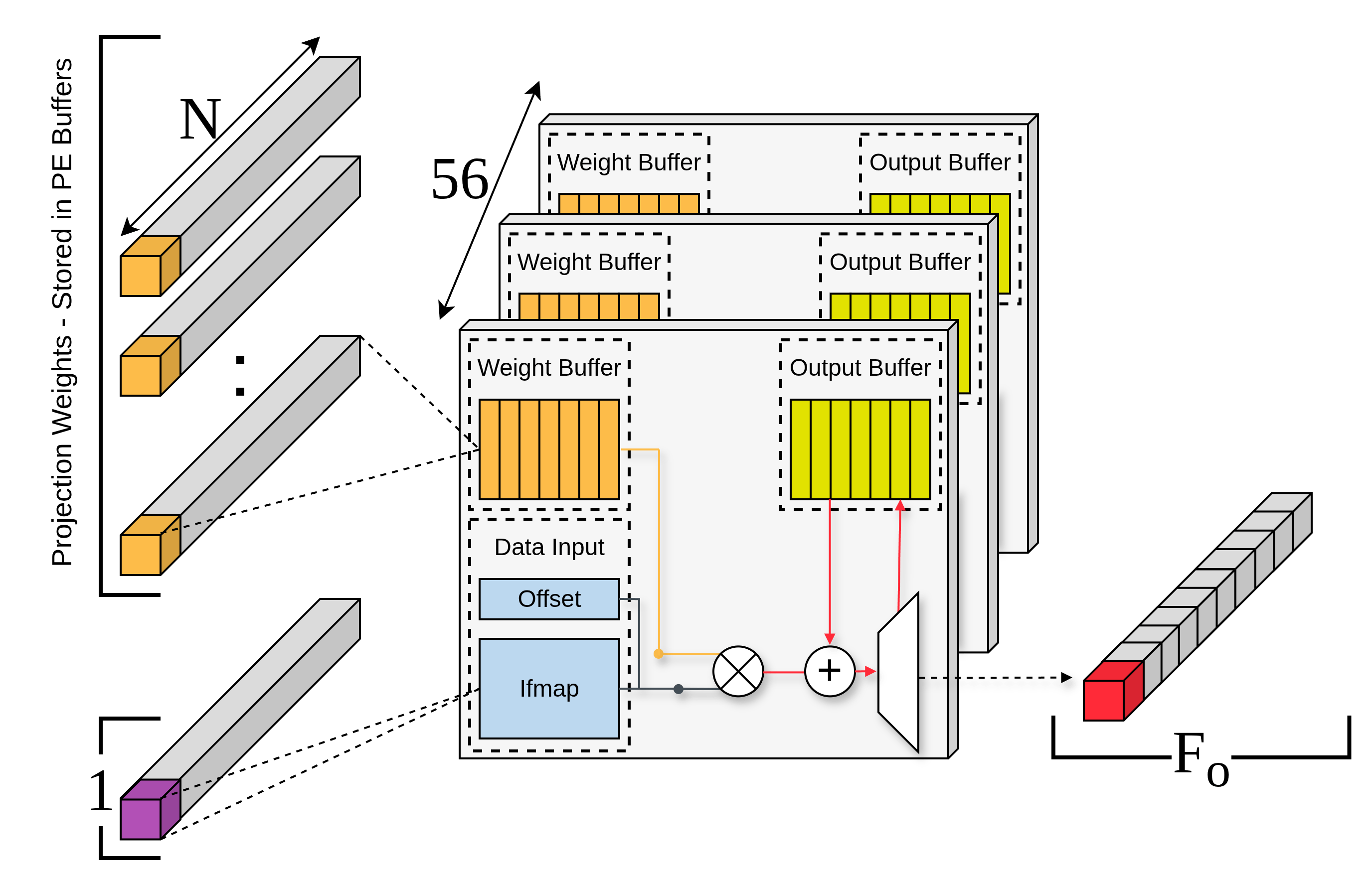} 
    \caption{Parallel architecture of the Projection Unit. Each of the 56 parallel engines holds a unique projection filter in its local Weight Buffer and accumulates partial sums for a single output channel in its Output Buffer.}
    \label{fig:projection_engine}
\end{figure}

The Projection Unit forms the final computational stage of the fused pipeline, performing the concluding $1{\times}1$ pointwise convolution that reduces, or “projects,” the expanded feature map channels from $M$ back to the desired output depth $N$. As illustrated in Figure~\ref{fig:projection_engine}, this unit is designed for maximum channel-level parallelism and consists of 56 parallel Projection Engines operating concurrently. Each engine is responsible for computing one complete output channel for a single pixel and contains its own private weight buffer implemented using distributed on-chip memory (LUTRAM). This memory organization eliminates access contention and routing overhead associated with a centralized weight buffer, allowing fully independent computation across all engines. The Projection Unit adopts an Output-Stationary (OS) dataflow. For each output pixel, partial sums for all 56 output channels remain stationary in their respective engine accumulators (the “Output Buffer” in Figure~\ref{fig:projection_engine}). In each iteration of the computation loop, a single 8-bit value from the Depthwise Unit’s post-processing pipeline is broadcast to all Projection Engines simultaneously. Each engine multiplies this shared input value with the corresponding weight from its local buffer and adds the result to its internal 32-bit accumulator. This process repeats for all $M$ input channels until the output for all 56 channels is fully accumulated. A dedicated post-processing pipeline then applies bias addition and requantization to produce the final 8-bit quantized results, which are made available to the CPU through explicit read instructions for subsequent software-level processing.
%-----------------------------------------------------------

%-----------------------------------------------------------

%-----------------------------------------------------------
\subsection{Pipeline Evolution and Timing Optimization}
%-----------------------------------------------------------
The evolution of the pipeline architecture reflects a series of deliberate design optimizations aimed at improving utilization and throughput without increasing hardware complexity. The initial version (v1) of the accelerator followed a fully sequential execution model, processing one output pixel to completion before beginning the next. In this configuration, each computational stage along with the final quantization and readback operations, executed in strict sequence for every pixel. Although straightforward to implement, this approach left most hardware resources idle for large portions of time, as downstream units could not begin work until the previous stage had finished. As a result, throughput was limited by the latency of the slowest stage.

To overcome this inefficiency, the second version (v2) introduced a three-stage \textit{inter-stage pipeline} that allowed the Expansion, Depthwise, and Projection units to operate concurrently on different pixels. By treating each of these units as an independent pipeline stage, the design enabled overlapping execution—for example, while the Depthwise Unit processed pixel $i$, the Expansion Unit computed pixel $i{+}1$, and the Projection Unit completed pixel $i{-}1$. This stage-level pipelining reduced idle cycles and maintained a steady flow of data across the accelerator, yielding a substantial improvement in throughput. The decision to pipeline at this level stemmed from the modular nature of the DSC structure, where each unit already performs a distinct and self-contained computation.

The final version (v3) extended this concept into a deeper, five-stage \textit{intra-stage pipeline}, achieving finer-grained parallelism by exploiting timing asymmetry within the compute stages themselves. Rather than simply dividing operations, the design separates the compute (MAC) and post-processing (Quantize) functions within the Expansion and Depthwise units—each with distinct latency and data dependencies. This creates five balanced pipeline stages: (1) Expansion MAC, (2) Expansion Quantize, (3) Depthwise MAC, (4) Depthwise Quantize, and (5) Projection MAC. With more pixels “in flight” simultaneously, this version maintains continuous utilization across all engines while avoiding unnecessary buffering or scheduling overhead. Overall, the progression from v1 to v3 transforms the accelerator from a sequential processor into a deeply pipelined system that aligns compute and dataflow timing for maximum efficiency.
%-----------------------------------------------------------
\begin{figure}[htbp]
    \centering
    \subfloat[Sequential Architecture]{
        \includegraphics[width=0.9\columnwidth]{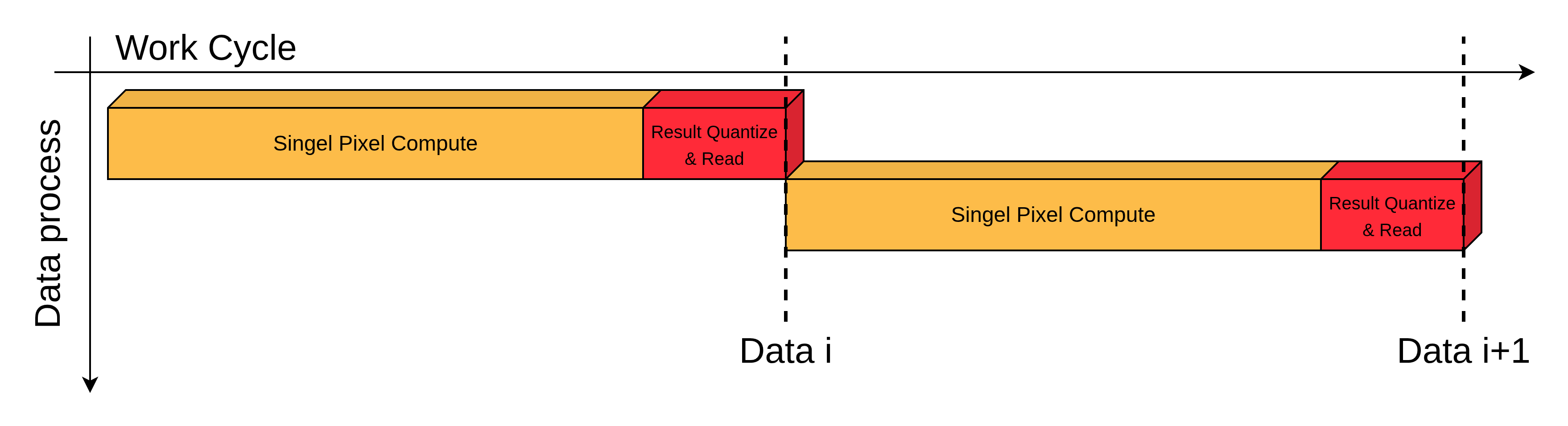}
        \label{fig:sequential_v1}
    }
    \\ % This creates the vertical break
    \subfloat[Inter-Stage Pipeline]{
        \includegraphics[width=0.8\columnwidth]{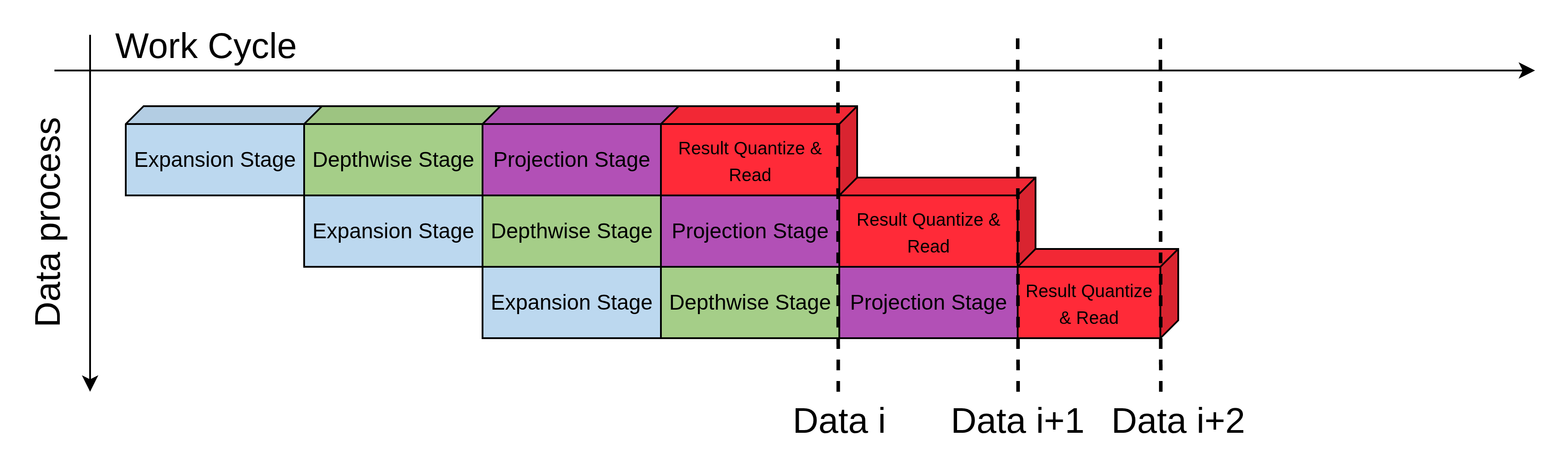}
        \label{fig:3_stage_pipeline_v2}
    }
    \\ % This creates the vertical break
    \subfloat[Intra-Stage Pipeline]{
        \includegraphics[width=0.8\columnwidth]{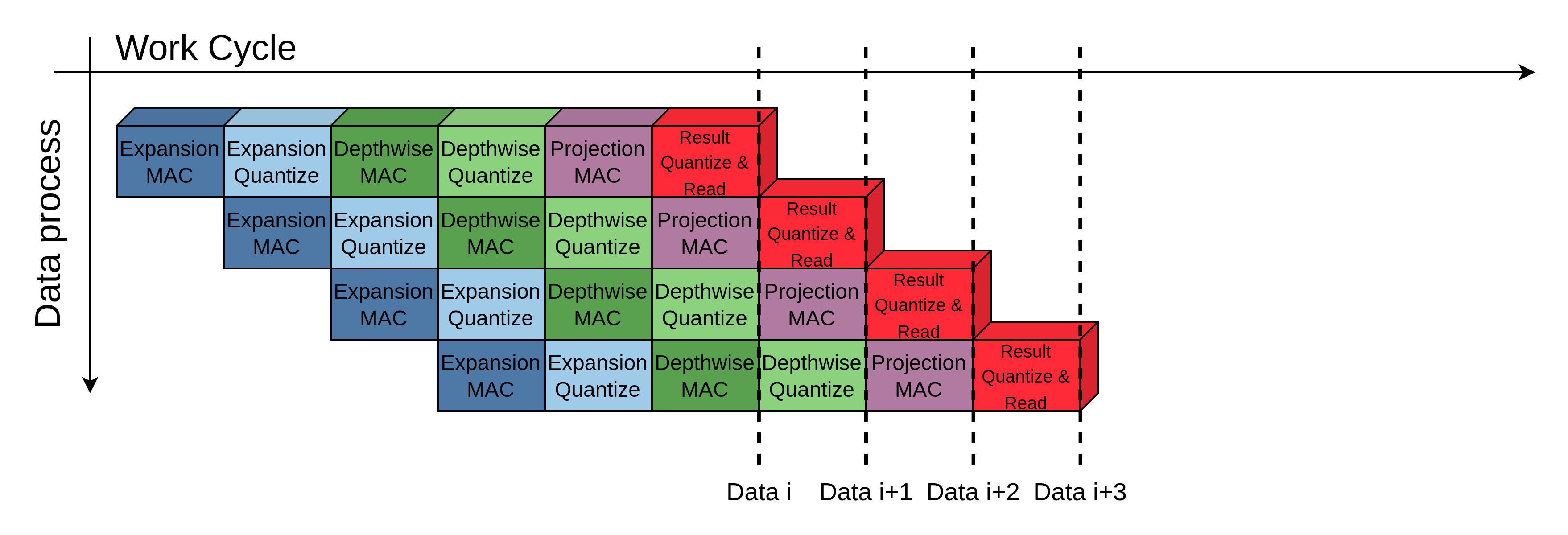}
        \label{fig:5_stage_pipeline_v3}
    }
    \caption{
    Evolution of pipeline architectures.
    (a) \textbf{Sequential (v1):} The baseline design processes one pixel completely before starting the next, leaving hardware resources idle between computations. 
    (b) \textbf{Inter-Stage Pipeline (v2):} Overlaps the Expansion, Depthwise, and Projection stages so the accelerator can process three different pixels concurrently. 
    (c) \textbf{Intra-Stage Pipeline (v3):} Further deepens the pipeline by splitting each stage into separate compute (MAC) and post-processing paths, enabling finer-grained parallelism and higher throughput.
    }
    \label{fig:pipeline_all}
\end{figure}
%-----------------------------------------------------------
\subsection{Heterogeneous On-Chip Memory Architecture}
%-----------------------------------------------------------

The primary goal of the memory subsystem is to eliminate the bandwidth and storage bottleneck caused by repeatedly storing and retrieving intermediate feature maps between the DSC stages. This is achieved through a heterogeneous on-chip memory design, where each stage employs a specialized buffer architecture tailored to its specific data access pattern. The Expansion and Depthwise stages use centralized BRAM-based buffers that provide high-capacity storage and efficient multi-port access for streaming data. In contrast, the Projection stage adopts a distributed LUTRAM-based memory organization, which offers low latency and localized data access for its highly parallel processing structure. This heterogeneous approach ensures that each stage achieves optimal throughput and minimal data-transfer overhead without the need for large, shared intermediate buffers.
%-----------------------------------------------------------

\begin{figure}[htbp]
	\centering
	\includegraphics[width=0.4\textwidth]{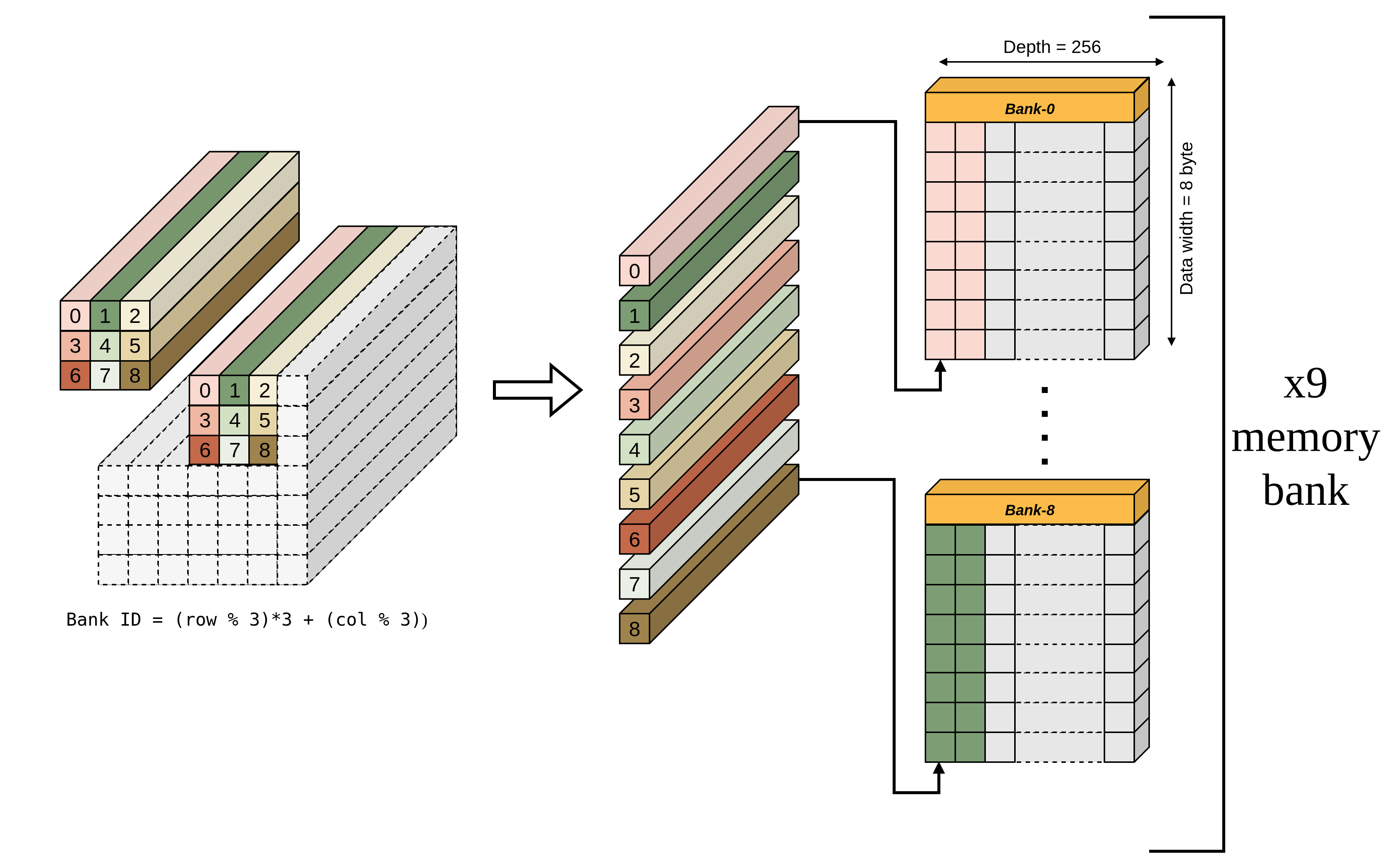}
	\caption{Banked memory architecture of the Ifmap Buffer for parallel window access. A 3x3 spatial window from the input feature map is mapped to nine parallel memory banks. As shown by the formula, the specific bank for each pixel is determined by its row and column position within the window. This structure allows all nine data words for a full 3x3 window to be read simultaneously in a single operation, enabling the high-throughput computation in the Expansion stage.}
	\label{fig:ifmap_memory}
\end{figure}

The IFMAP Buffer is designed to store the input feature map and provide simultaneous access to all nine pixels of a $3{\times}3$ spatial window required by the parallel Expansion Engines. As shown in Figure~\ref{fig:ifmap_memory}, the buffer adopts a nine-bank parallel memory structure, implemented using nine independent BRAM blocks. A pixel located at coordinates $(\text{row}, \text{col})$ is mapped to a specific memory bank according to the rule:
\[
\text{Bank ID} = (\text{row} \bmod 3) \times 3 + (\text{col} \bmod 3)
\]
This banking scheme guarantees that every $3{\times}3$ window of pixels is distributed across nine different banks. As a result, all nine data words corresponding to one spatial window can be read simultaneously in a single clock cycle. This organization provides the high data bandwidth required to sustain the fully parallel operation of the Expansion Unit, ensuring that all nine Expansion Engines remain continuously fed without stalling.
%-----------------------------------------------------------

%-----------------------------------------------------------

The Expansion Filter Buffer is implemented as a single, large BRAM block that stores all $M$ expansion filters sequentially in memory. Each filter corresponds to a $1{\times}1{\times}N$ kernel, where $N$ is the input channel depth and is always a multiple of eight. During computation, the buffer outputs one 8-channel (64-bit) chunk of filter data per cycle, which is broadcast simultaneously to all nine parallel Expansion Engines. This organization ensures that all engines operate on the same output channel at a given time, enabling synchronized processing while minimizing memory access overhead. By streaming the filter data in compact 8-byte words and leveraging a broadcast mechanism, the design achieves high bandwidth efficiency and reduces redundant memory reads.
%-----------------------------------------------------------

\begin{figure}[htbp]
	\centering
	\includegraphics[width=0.4\textwidth]{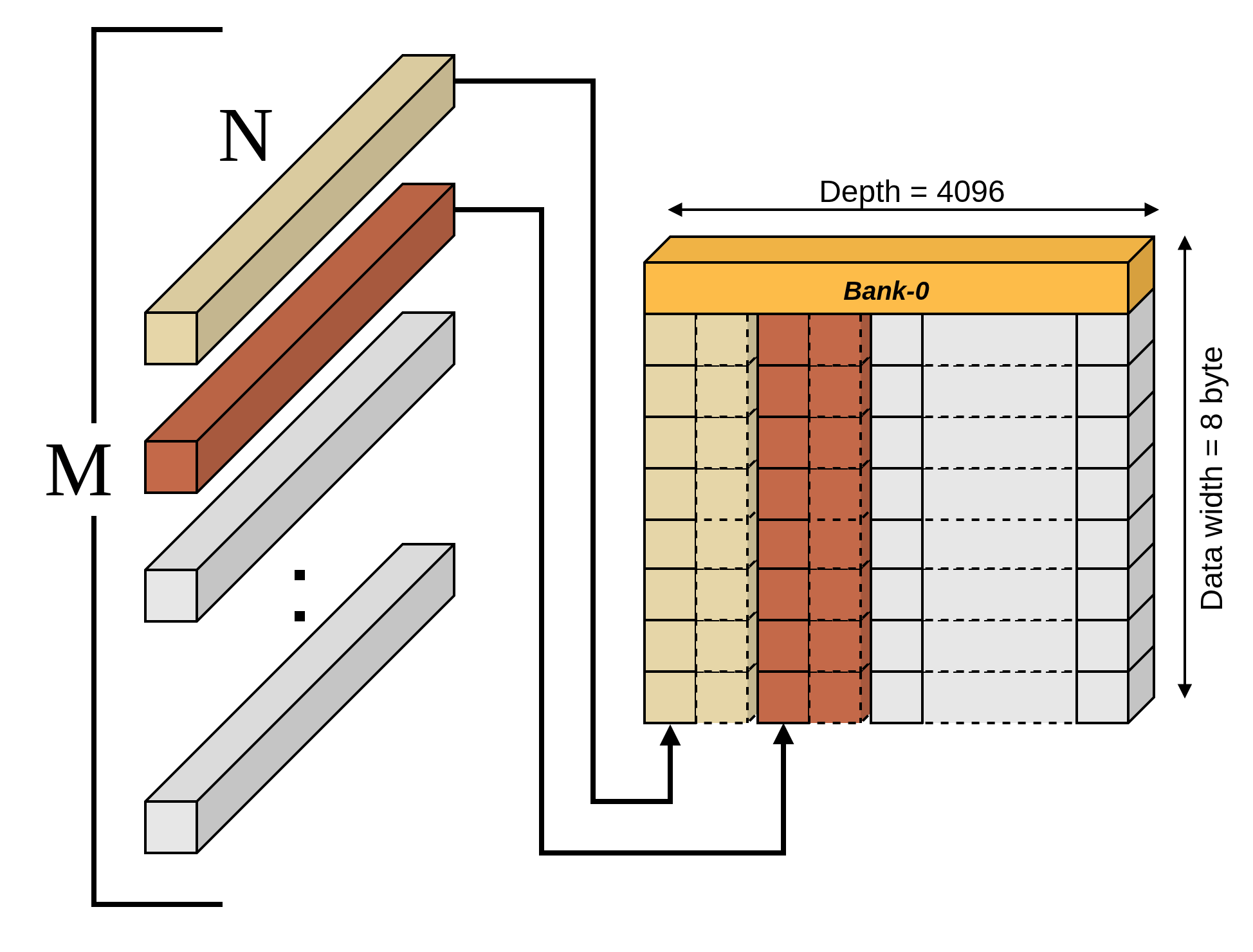}
	\caption{Memory Architecture of the Expansion Filter Buffer. The M logical 1x1xN filters are stored sequentially. For a given computation, a filter is read in N/8 sequential accesses, where N is the input channel depth and is a multiple of 8. Each 8-byte (64-bit) word fetched from this buffer is broadcast to all nine parallel Expansion Engines. }
	\label{fig:ex_filter_buffer}
\end{figure}
%-----------------------------------------------------------

The Depthwise Filter Buffer adopts a nine-bank parallel memory organization, mirroring the structure of the IFMAP buffer. Each of the nine banks corresponds to a fixed spatial position within the $3{\times}3$ depthwise kernel(e.g., Bank~0 stores the top-left weight across all filters, Bank~1 stores the top-middle weight, and so on). This arrangement ensures that for any of the $M$ depthwise filters in a layer, all nine weights can be fetched simultaneously in a single clock cycle. The resulting 72-bit word, representing the complete $3{\times}3$ filter, is then supplied directly to the Depthwise Engine. This parallel access pattern perfectly aligns with the engine’s nine-way MAC structure, providing one weight per multiplier and enabling single-cycle filter loading for continuous computation.
%-----------------------------------------------------------

\begin{figure}[htbp]
	\centering
	\includegraphics[width=0.4\textwidth]{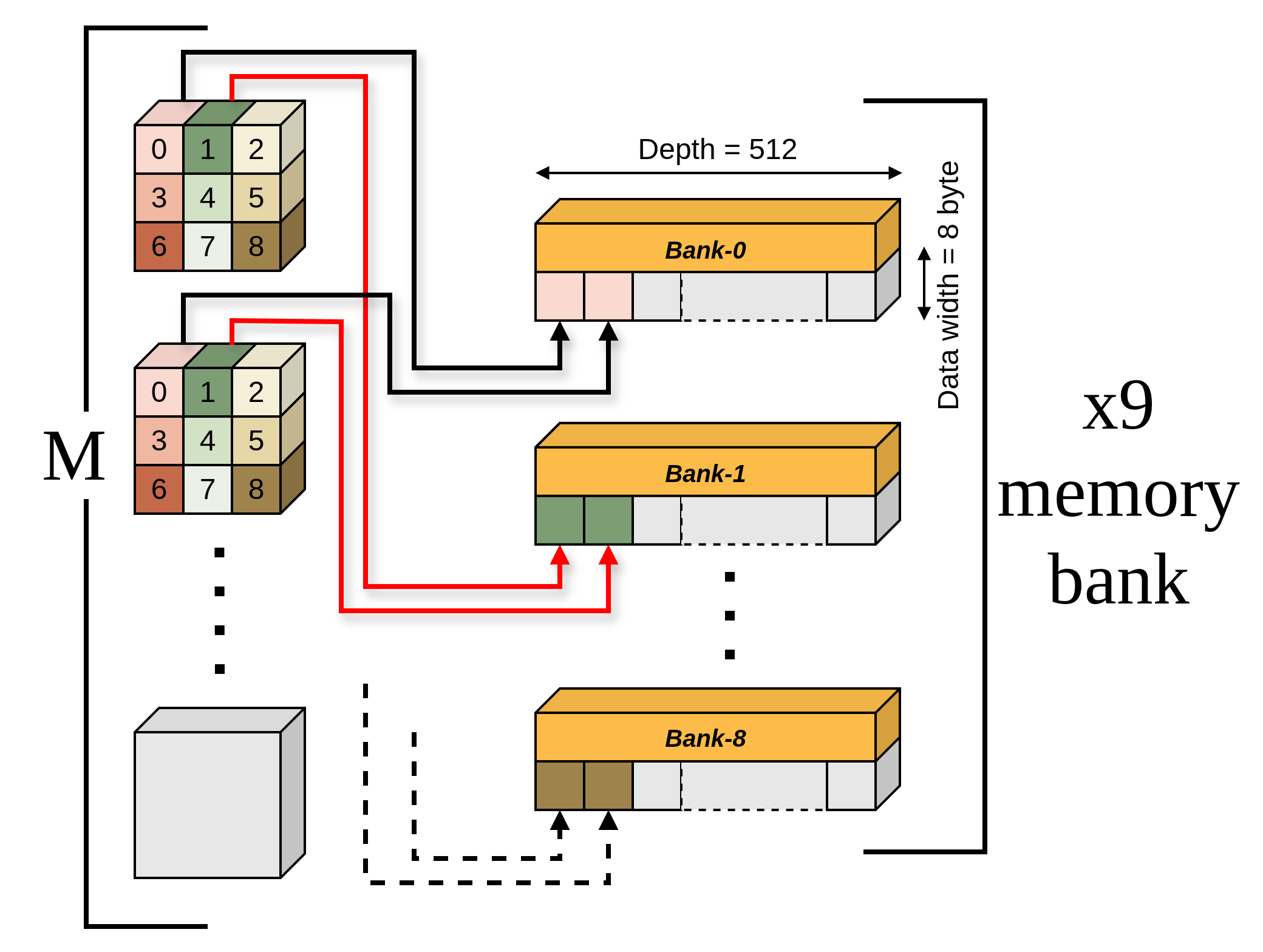}
	\caption{Banked memory architecture of the Depthwise Filter Buffer. To enable parallel processing, each of the nine weights in a 3x3 depthwise filter is stored in its own dedicated memory bank. All M filters for a given layer are stored in this way, allowing the hardware to read an entire 3x3 (72-bit) filter in a single clock cycle.}
	\label{fig:dw_filter_buffer}
\end{figure}
%-----------------------------------------------------------

Unlike the BRAM-based buffers used in earlier stages, the Projection stage employs a distributed memory architecture. Each of the 56 parallel Projection Engines includes its own small, private weight buffer that stores the coefficients of its unique $1{\times}1$ filter. These buffers are implemented using LUTRAM (distributed on-chip memory) rather than shared BRAM, providing ultra-low-latency access and eliminating memory port contention. This localized memory approach is well-suited for highly parallel designs, as it minimizes routing complexity and ensures that each engine can independently access its weights without interference.
%-----------------------------------------------------------
\subsection{On-the-Fly Padding Logic}
%-----------------------------------------------------------

\begin{figure}[htbp]
    \centering
    \subfloat[Conventional Method with Explicit Padding]{
        \includegraphics[width=0.9\columnwidth]{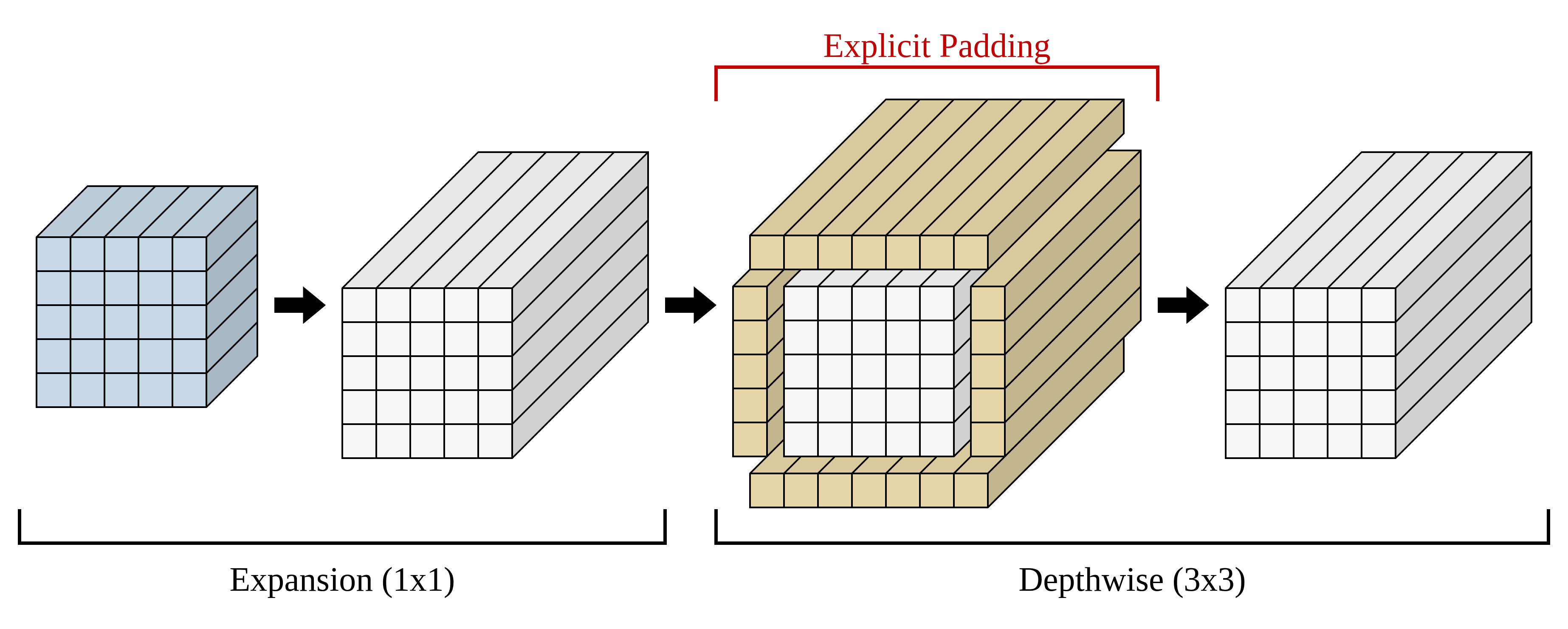}
        \label{fig:real_padding}
    }
    \\ % This creates the vertical break
    \subfloat[Proposed Method with On-the-Fly Padding]{
        \includegraphics[width=0.8\columnwidth]{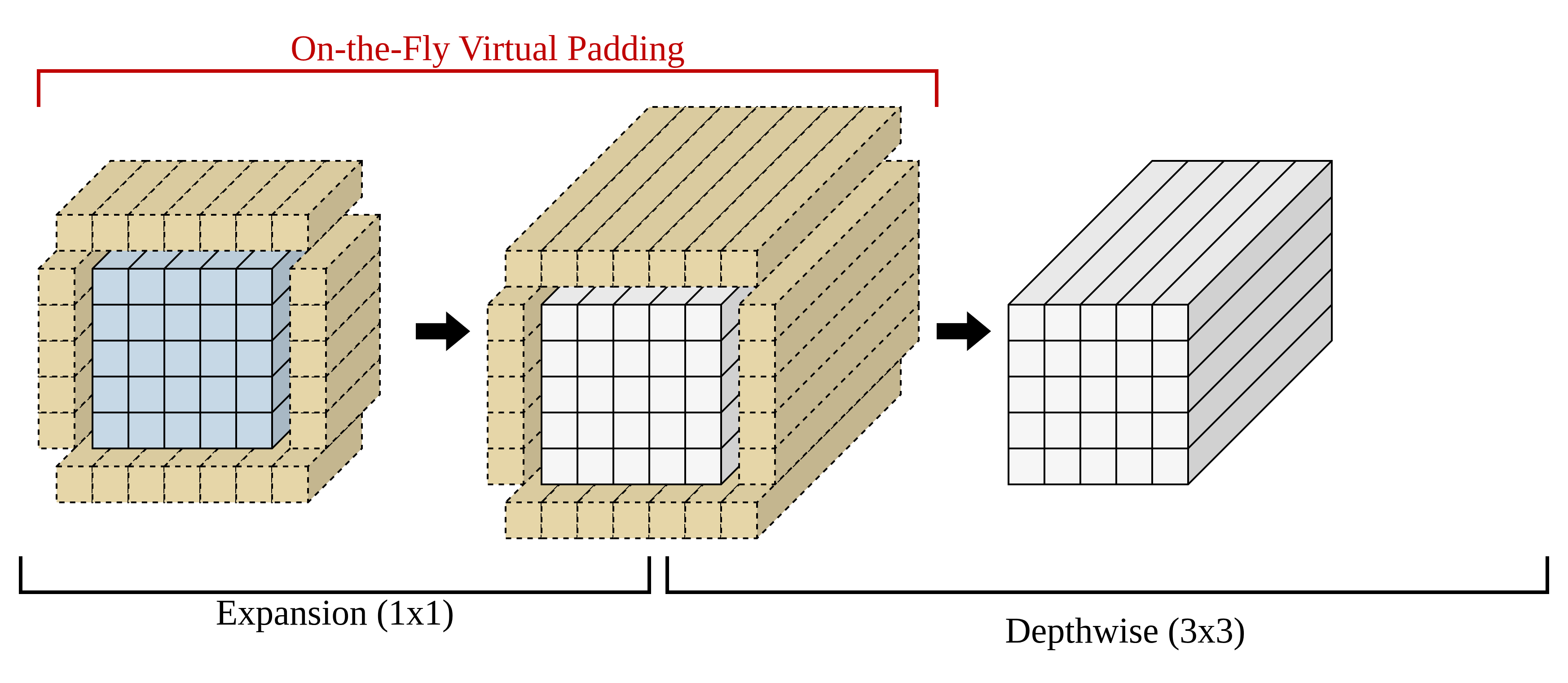}
        \label{fig:virtual_padding}
    }
     \caption{Comparison of padding methodologies.   \textbf{(a)} The conventional approach requires an explicit padding step, creating a larger intermediate tensor that must be physically stored in an on-chip buffer.   \textbf{(b)} Our proposed method uses on-the-fly hardware logic to handle padding virtually. The intermediate padded tensor is never stored, eliminating the buffer and reducing memory traffic.}
    \label{fig:padding_logic_comparison}
\end{figure}

In a conventional layer-by-layer implementation, padding is handled in software by explicitly expanding the intermediate feature map with zeros before convolution, as shown in Figure~\ref{fig:real_padding}. However, in our fused architecture, the intermediate feature map ($F_1$) is never fully stored, making traditional padding infeasible. To address this, the Ifmap Buffer integrates hardware-managed on-the-fly padding logic that operates during each read cycle. When the Expansion or Depthwise unit requests a $3{\times}3$ window, the address generation logic checks whether any of the accessed pixel coordinates fall outside the valid feature map boundary. For valid coordinates, data are fetched normally from BRAM, whereas out-of-bound accesses are automatically replaced by the quantization zero-point value. As illustrated in Figure~\ref{fig:virtual_padding}, this method eliminates the need to allocate memory for padded tensors and completely removes software padding overhead, enabling seamless fused execution and substantial on-chip memory savings.
%-----------------------------------------------------------

%-----------------------------------------------------------
\section{Experimental Results}
%-----------------------------------------------------------
\subsection{Overview and Methodology}
%-----------------------------------------------------------
To evaluate the proposed fused DSC accelerator, we conducted a comprehensive set of experiments across both FPGA and ASIC design domains. Four primary configurations were evaluated: a software-only baseline (v0) running on the base VexRiscv-Litex SoC \cite{spinalhdl_vexriscv_2023}, and our three accelerator iterations (v1, v2, and v3). All accelerator versions were integrated into this SoC using the CFU-Playground framework \cite{prakash2023cfu}, which runs on the VexRiscv RISC-V\cite{spinalhdl_vexriscv_2023} processor to ensure a consistent software and hardware environment. The tests were performed using the TensorFlow Lite (TFLite) implementation of the MobileNetV2 model, executed layer by layer to measure accelerator performance and correctness. Synthesis and implementation on FPGA were carried out using Xilinx Vivado targeting the Digilent Nexys A7-100T board which is equipped with an Artix-7 XC7A100T device. The board's available resources are summarized in Table~\ref{tab:fpga_resources}. This platform provides a practical validation of functionality, performance, and throughput improvement under realistic TinyML hardware constraints.

To further verify scalability and energy efficiency beyond the FPGA fabric, we synthesized the accelerator at the ASIC level using Cadence Genus\cite{cadence_genus_2024} for logic synthesis and the CACTI\cite{muralimanohar2009cacti} tool for memory modeling and power estimation. While Vivado results demonstrate real-time operation, the Genus and CACTI analyses provide deeper insight into achievable frequency, area, and power under modern semiconductor technologies (40~nm and 28~nm nodes). This dual-domain evaluation allows us to assess both the functional performance on resource-limited hardware and the potential efficiency gains in ASIC-class integration, giving a holistic view of the proposed accelerator’s practicality and design maturity.

%-----------------------------------------------------------
\begin{table}[!t]
\renewcommand{\arraystretch}{1.3}
\caption{Available Hardware Resources on the Artix-7 XC7A100T FPGA}
\label{tab:fpga_resources}
\centering
\begin{tabular}{c c c c }
\hline
\textbf{LUTs} & \textbf{FFs} & \textbf{DSPs} & \textbf{Block RAM (36Kb)}  \\
\hline
63400 & 126800 & 240 & 135  \\
\hline
\end{tabular}
\end{table}

%-----------------------------------------------------------
\subsection{FPGA Performance: Pipeline Evolution and Speedup}
\begin{figure}[htbp]
    \centering
    \includegraphics[width=0.9\columnwidth]{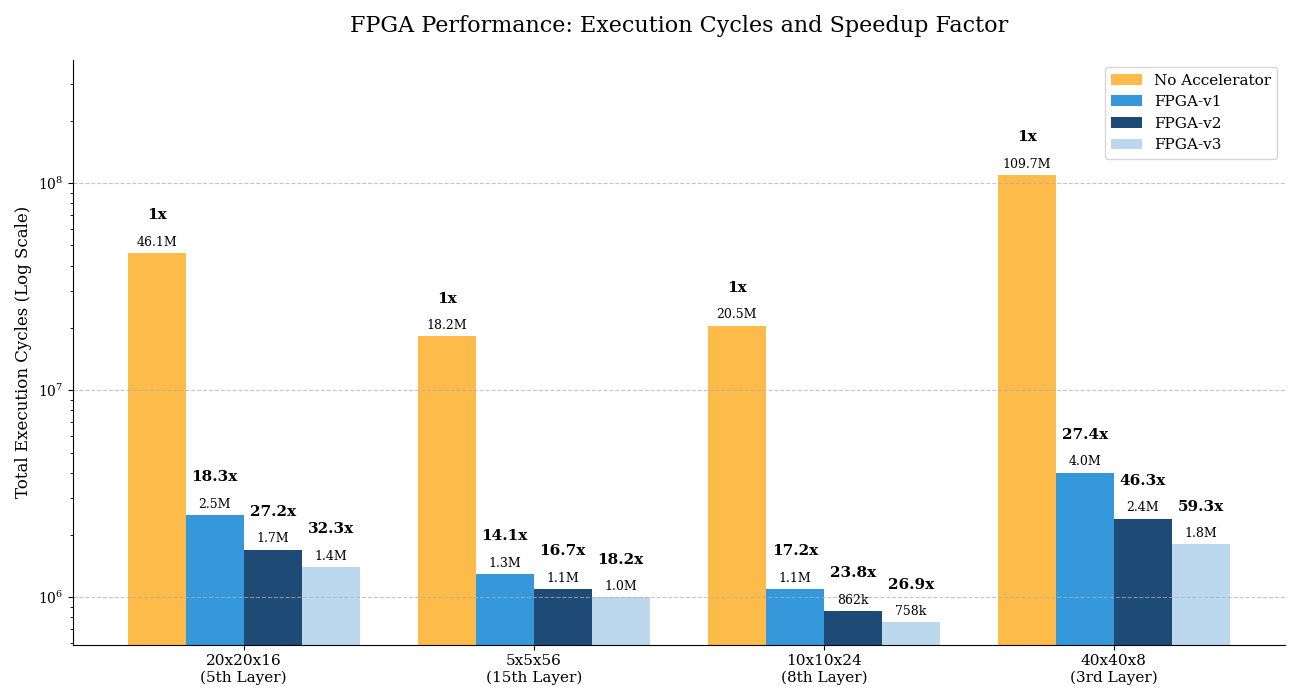}
    \caption{FPGA performance comparison across four different MobileNetV2 bottleneck layers. The bars represent the total execution cycles on a logarithmic scale for the unaccelerated baseline and three accelerator versions. The labels indicate the speedup factor relative to the baseline, demonstrating significant performance gains and consistent improvement from v1 to v3.}
    \label{fig:speedups}
\end{figure}

%-----------------------------------------------------------
To assess the impact of the proposed pipeline refinements, four accelerator configurations were evaluated on the Artix-7 FPGA at a 100~MHz system clock. Figure~\ref{fig:speedups} illustrates the performance gains from our iterative architectural evolution. The initial sequential design (v1) already provides a significant speedup, such as 27.4x on the 3th layer. Introducing the 3-stage inter-stage pipeline (v2) further improves performance by overlapping the main compute stages, increasing the speedup to 46.3x on the same layer. The final 5-stage intra-stage pipeline (v3) achieves the highest throughput by minimizing hardware idle time, resulting in a maximum speedup of \textbf{59.3x} for the 3rd layer. This progressive enhancement demonstrates the effectiveness of pipelining at both the stage and sub-stage levels in maximizing resource utilization and reducing overall execution time.

%-----------------------------------------------------------

%-----------------------------------------------------------
Table~\ref{tab:fpga_utilization} details the hardware cost for this performance evolution. Crucially, the logic, BRAM, and DSP resources are identical across all three accelerator versions (v1, v2, and v3). This confirms that the 2.4x (v1 to v2) and 3.2x (v1 to v3) speedups stem purely from architectural restructuring---specifically, adding registers to enable inter- and intra-stage pipelining---and not from adding more computational hardware. The v3 design shows a slight power reduction compared to v1 and v2, which we attribute to more efficient clock gating and reduced signal toggling in the deeply pipelined, continuously active datapath.

%-----------------------------------------------------------
\begin{table}[!h]
\renewcommand{\arraystretch}{1.3}
\caption{FPGA resource utilization and power comparison (Vivado @ 100 MHz)}
\label{tab:fpga_utilization}
\centering
\begin{tabular}{l c c c c}
\hline
\textbf{Resource} & \textbf{Base\cite{spinalhdl_vexriscv_2023}} & \textbf{FPGA-v1} & \textbf{FPGA-v2} & \textbf{FPGA-v3} \\
\hline
LUTs & 4,438 & 20,922 & 20,922 & 20,922 \\
Flip-Flops (FF) & 3,804 & 17,752 & 17,752 & 17,752 \\
BRAM (36Kb) & 15 & 97 & 97 & 97 \\
DSP Slices & 5 & 178 & 178 & 178 \\
\hline
Power (W) & 0.673 & 1.275 & 1.303 & 1.121 \\
\hline
\end{tabular}
\end{table}

%-----------------------------------------------------------
Table~\ref{tab:cfu_comparison} provides a detailed view of both performance and hardware cost, comparing our final v3 design against the software-only baseline (v0) and the original CFU-Playground accelerator \cite{prakash2023cfu}. The performance data in Table~\ref{tab:cfu_comparison}(A) highlights the magnitude of improvement: our v3 design completes a representative bottleneck layer in 1.8 million cycles, whereas the baseline software on the Litex-SoC takes 109.7 million cycles.

More significantly, our v3 accelerator is \textbf{20-30x faster} than the CFU-Playground accelerator on the same layers. Table~\ref{tab:cfu_comparison}(B) details the hardware trade-off for this gain. Our design is intentionally larger. This trade-off is at the core of our design: the CFU-Playground accelerator only targets 1x1 convolutions, leaving the 3x3 depthwise kernel and all inter-layer data movement to the CPU. In contrast, our architecture accelerates the entire \textbf{Ex$ \rightarrow$ Dw$ \rightarrow$ Pr} block, eliminating the CPU data-shuffling bottleneck and solving the intermediate memory wall, which accounts for its substantially higher performance.

%-----------------------------------------------------------
% --- Comparison Table (Cycles + Resources) ---
\begin{table}[!h]
\renewcommand{\arraystretch}{1.3}
\caption{FPGA Performance and Resource Comparison vs.\ CFU-Playground\cite{prakash2023cfu}}
\label{tab:cfu_comparison}
\centering

% --- (A) Performance Table ---
\resizebox{\columnwidth}{!}{%
\begin{tabular}{l c c c}
\hline
\multicolumn{4}{c}{\textbf{(A) Performance (Total Cycles @ 100 MHz)}} \\
\hline
\textbf{Layer (Workload)} & \textbf{Baseline\cite{spinalhdl_vexriscv_2023}} & \textbf{CFU-Playground \cite{prakash2023cfu}} & \textbf{Our FPGA-v3} \\
\hline
3rd Layer & 109.7M & 45.6M & \textbf{1.8M} \\
5th Layer & 46.1M & 32.7M & \textbf{1.4M} \\
8th Layer & 20.5M & 8.4M & \textbf{0.76M} \\
15th Layer & 18.2M & 5.4M & \textbf{1.0M} \\
\hline
\end{tabular}}

\vspace{0.5em}

% --- (B) Resource Table ---
\resizebox{\columnwidth}{!}{%
\begin{tabular}{l c c c}
\hline
\multicolumn{4}{c}{\textbf{(B) Resource Utilization}} \\
\hline
\textbf{Resource} & \textbf{Baseline\cite{spinalhdl_vexriscv_2023}} & \textbf{CFU-Playground \cite{prakash2023cfu}} & \textbf{Our FPGA-v3} \\
\hline
LUTs & 4,438 & 6,055 & \textbf{20,922} \\
FFs & 3,804 & 4,501 & \textbf{17,752} \\
BRAM (36Kb) & 15 & 24 & \textbf{97} \\
DSP Slices & 5 & 18 & \textbf{178} \\
\hline
\end{tabular}}
\\[2pt]
\parbox{\columnwidth}{\footnotesize \textit{**Baseline (v0)} refers to the base VexRiscv-Litex SoC with an empty CFU interface, running software-only kernels. \textit{CFU-Playground} refers to the 1x1 convolution accelerator from Prakash et al. \cite{prakash2023cfu}.}
\end{table}

%-----------------------------------------------------------
For a fair assessment against other accelerators in the same environment, Table~\ref{tab:cfu_mnv2_comparison} summarizes recent CFU-Playground works targeting the \textit{MobileNetV2 TFLite} model. These accelerators show different optimization goals. Lightweight approaches like \textbf{Sabih et al. \cite{sabih2024hardware}} focus on sparsity, adding minimal hardware (~108 LUTs, 2 DSPs) to skip zero-value computations for a ~5.1x model speedup; power was not reported for this design. The original CFU-Playground reference from \textbf{Prakash et al. \cite{prakash2023cfu}} (\texttt{mnv2\_first}) provides a 2.4x speedup over the CPU baseline, consuming 0.742 W. The general-purpose \textbf{AI-ISP accelerator \cite{wu2024ai}} achieved a 15.8x speedup over the \textit{total model} runtime of the Prakash et al. accelerator, at a cost of 6,308 LUTs and 138 DSPs, while consuming 1.58 W.

Our work, in contrast, focuses on the memory bottleneck. As shown in Table~\ref{tab:cfu_mnv2_comparison}, our fused-pipeline design achieves a \textbf{59.3x speedup over the CPU baseline} for the 3rd layer. This is \textbf{25.3x faster than the original Prakash et al. \cite{prakash2023cfu} accelerator} on the same layer. This large performance gain is the direct result of our specialized design (16,484 LUTs, 173 DSPs), which is the only one to implement a zero-buffer fused dataflow for the full Ex$\rightarrow$Dw$\rightarrow$Pr block. Notably, despite its significantly higher resource usage and performance, our accelerator consumes \textbf{1.12 W}, which is 29\% less power than the AI-ISP general-purpose engine. This comparison confirms that our full-fusion approach provides a much larger performance gain by solving the system-level dataflow problem, and it does so with high power efficiency.

%-----------------------------------------------------------
\begin{table}[!h]
\renewcommand{\arraystretch}{1.25}
\caption{Comparison of CFU-Playground–based MobileNetV2 Accelerators}
\label{tab:cfu_mnv2_comparison}
\centering
\resizebox{\columnwidth}{!}{%
\begin{tabular}{l c c c}
\hline
\textbf{Work} & \textbf{Speedup Metric} & \textbf{Reference Baseline} & \textbf{Power (W)} \\
\hline
\textbf{This Work (v3)} & \textbf{59.3×} / \textbf{25.3×} & CPU / Prakash~\cite{prakash2023cfu} & \textbf{1.12} \\
Wu \emph{et al.}~\cite{wu2024ai} & 15.8× & Prakash~\cite{prakash2023cfu} & 1.58 \\
Sabih \emph{et al.}~\cite{sabih2024hardware} & $\sim$5.1× & CPU baseline & N/A \\
Prakash \emph{et al.}~\cite{prakash2023cfu} & $\sim$2.4× & CPU baseline & 0.742 \\
\hline
\end{tabular}}
\\[3pt]
\parbox{\columnwidth}{\footnotesize
\textit{Notes:} Reported speedups correspond to \textbf{MobileNetV2} inference unless noted.
This Work and Prakash~\cite{prakash2023cfu} report speedups for the \textit{3rd bottleneck layer};
Wu~\cite{wu2024ai} reports model-level performance; 
Sabih~\cite{sabih2024hardware} targets MobileNetV2–VWW.}
\end{table}

%-----------------------------------------------------------
Despite the increase in resource usage compared with the minimal CFU-Playground design, our accelerator remains compact and efficient, consuming only 33\% of the Artix-7's LUTs and 74\% of its DSPs. All pipeline versions sustained a stable 100~MHz operating frequency, confirming that the added pipeline depth did not create new timing bottlenecks. Overall, these results demonstrate that substantial throughput improvement—up to 59x—is achievable on a low-cost FPGA through architectural parallelism and dataflow fusion, without requiring increases in clock frequency.

%-----------------------------------------------------------
\subsection{ASIC Synthesis and Power-Area Analysis}
%-----------------------------------------------------------
To evaluate the scalability and efficiency of the proposed architecture beyond FPGA implementation, we synthesized the accelerator at the ASIC level using Cadence Genus for logic synthesis and CACTI for memory modeling. These tools provide realistic estimates of area and power consumption at modern technology nodes, allowing a fair assessment of the design’s potential for integration into low-power embedded SoCs.

The logic part of the accelerator—including compute engines, pipelines, and control units—was synthesized using a standard-cell library for both 40\,nm and 28\,nm CMOS processes. The SRAM buffers were modeled in CACTI 7.0 to capture realistic density and power characteristics of small on-chip memories used in TinyML systems. The operating frequencies were set to 300\,MHz for 40\,nm and 2\,GHz for 28\,nm to represent typical voltage–frequency targets at each node.

\begin{table}[!h]
\renewcommand{\arraystretch}{1.3}
\caption{Total Accelerator Area and Power Summary at 40\,nm and 28\,nm Nodes}
\label{tab:asic_summary}
\centering
\begin{tabular}{lcc}
\hline
\textbf{Metric} & \textbf{40\,nm @ 300\,MHz} & \textbf{28\,nm @ 2\,GHz} \\
\hline
Logic Area (mm$^2$) & 0.976 & 0.284 \\
Memory Area (mm$^2$) & 0.218 & 0.072 \\
\textbf{Total Area (mm$^2$)} & \textbf{1.194} & \textbf{0.356} \\
\hline
Logic Power (mW) & 145.7 & 821.8 \\
Memory Power (mW) & 106.5 & 88.2 \\
\textbf{Total Power (mW)} & \textbf{252.2} & \textbf{910.0} \\
\hline
\end{tabular}
\\[2pt]
\parbox{\columnwidth}{\footnotesize \textit{*Area and power results reflect the accelerator CFU only and do not include the VexRiscv CPU core}.}
\end{table}

The synthesis results confirm that the proposed fused architecture is compact and scalable. At the 40\,nm node, the complete accelerator occupies only 1.19\,mm$^2$ and consumes approximately 252\,mW at 300\,MHz,which falls well within the power budget of TinyML edge devices that typically operate in the sub-watt range. When scaled to 28\,nm, the design achieves a threefold area reduction to 0.36\,mm$^2$ while maintaining under 1\,W total power even at a 2\,GHz frequency target, demonstrating strong energy efficiency and frequency scalability.

Importantly, the logic-to-memory power ratio remains balanced across both nodes, confirming that the fused, zero-buffer dataflow effectively limits memory energy consumption—a key challenge in TinyML accelerators.  These results validate that the proposed architecture not only improves computational throughput on FPGA,  but also maintains power and area efficiency when synthesized for ASIC implementation,  making it well-suited for future integration in resource-constrained TinyML SoCs.

%-----------------------------------------------------------
\subsection{Comparative Analysis of Memory Traffic Reduction}
%-----------------------------------------------------------
In TinyML and Edge-AI accelerators, overall efficiency is often limited by the amount of data moved between layers rather than by arithmetic operations. To highlight the impact of our fused pixel-wise design, we first analyze the baseline layer-by-layer (L-by-L) execution of MobileNetV2 on the FPGA system. Table~\ref{tab:memory_access_fpga} shows the total intermediate memory access cycles and data transferred for representative residual blocks. Each block requires millions of cycles to move intermediate feature maps between the Expansion, Depthwise, and Projection stages, clearly showing that memory access dominates total execution time.

%-----------------------------------------------------------
\begin{table}[!h]
\renewcommand{\arraystretch}{1.3}
\caption{Intermediate Memory Access Analysis for Baseline Execution}
\label{tab:memory_access_fpga}
\centering
\resizebox{\columnwidth}{!}{%
\begin{tabular}{l c c}
\hline
\textbf{Layer (Workload)} & \textbf{Intermediate Access Cycles} & \textbf{Data Moved (Bytes)} \\
\hline
40$\times$40$\times$8 (3rd)  & 14.0 M & 307{,}200 \\
20$\times$20$\times$16 (5th) & 7.6 M  & 153{,}600 \\
10$\times$10$\times$24 (8th) & 2.7 M  & 57{,}600 \\
5$\times$5$\times$56 (15th)  & 1.8 M  & 33{,}600 \\
\hline
\end{tabular}}
\end{table}

%-----------------------------------------------------------
Our fused architecture removes these intermediate reads and writes completely.  Only the input feature map and three filters (Ex, Dw, Pr) are read once, and the output feature map is written once.  This eliminates redundant memory traffic, resulting in about \textbf{87\%} total data-movement reduction compared with the baseline.  To understand how this compares to existing DSC accelerators, we review several related works that address the same memory wall challenge. 

%-----------------------------------------------------------
\begin{table*}[!t]
\renewcommand{\arraystretch}{1.3}
\caption{Memory optimization strategies and FPGA implementations for DSC acceleration.}
\label{tab:mem_reduction}
\centering
\resizebox{\textwidth}{!}{%
\begin{tabular}{lccccccc}
\hline
\textbf{Work} & \textbf{Platform} & \textbf{Model} & \textbf{Precision} & \textbf{Method} & \textbf{Intermed. Buffer} & \textbf{Resources (LUT/FF/BRAM)} & \textbf{Reduction} \\
\hline
\textbf{This work (v3)} & Nexys4 DDR (A7) & MNV2 & INT8 & \textbf{Zero-buffer fusion (Ex–Dw–Pr)} & \textbf{None} & \textbf{16.7k / 15.4k / 81.5} & \textbf{87\%} \\
RAMAN~\cite{krishna2024raman} & Efinix Ti60 & MNV1 & INT2/4/8 & Pruning + sparsity & Cache/GLB & 37.2k / 8.6k / 168 & 34.5\% \\
Lei~Xuan~\cite{xuan2022fpga} & VC709 (V7) & MNV2 & INT4 & Partial fusion (Dw$\rightarrow$Pr) & Row/Tile SRAM & 107k / 74.4k / 13.7\,Mb & 80.5\% \\
Zhiyuan~Zhao~\cite{zhao2025high} & ZC706 (Zynq) & MNV2 & INT8 & Hybrid multi-CE streaming & Hybrid SRAM & 163k / 189k / 329.5 & 83.4\% \\
Jixuan~Li~\cite{zhao2025high} & VC709 (V7) & MNV2 & INT8 & Double-layer MAC (Dw+Pr) & SRAM after PW1 & 65k / 60k / 308 & 41.34\% \\
\hline
\end{tabular}}
\begin{tablenotes}[flushleft]          
\footnotesize
\item[] \emph{Normalization.} Reductions are reported versus each work’s stated baseline: L-by-L (ours, Xuan, Li), single-CE off-chip baseline (Zhao), and non-sparse cache traffic (RAMAN). When only BRAM/DRAM access counts are provided, we derive bytes using 8-bit tensors unless a different precision is explicitly reported. Platform and resource figures are not directly comparable across vendors/families (e.g., Xilinx vs. Efinix).
\end{tablenotes}
\end{table*}

%-----------------------------------------------------------
While performance metrics such as GOPS or FPS are widely reported, they are not ideal for cross-architecture comparison.  Our accelerator is implemented as a RISC-V CFU tightly coupled to the CPU, not a standalone compute core,  so it includes CPU–CFU control overhead that pure accelerators do not.  Therefore, this section focuses on architectural effectiveness in reducing memory movement—one of the main energy and latency bottlenecks in TinyML devices.  

%-----------------------------------------------------------
Compared with prior designs that rely on partial fusion or hybrid buffering, our fused pixel-wise architecture removes all intermediate feature-map transfers entirely.This leads to the largest reduction in total bytes moved while staying within a compact resource footprint suitable for embedded FPGAs.In summary, this design demonstrates that full fusion of Ex–Dw–Pr within a tightly coupled RISC-V CFU can deliver competitive performance and energy efficiencyfor the TinyML domain without requiring large buffers or high-end hardware.

%-----------------------------------------------------------
 \section{Conclusion}
%-----------------------------------------------------------
This work presented a memory-efficient hardware accelerator for Depthwise Separable Convolutions (DSCs) targeting the TinyML domain. We identified the intermediate feature map buffering between the Expansion, Depthwise, and Projection stages as the primary memory bottleneck in conventional layer-by-layer designs. To address this challenge, we introduced a fused, pixel-wise dataflow that eliminates intermediate memory transfers entirely by streaming data directly between convolution stages.

The proposed architecture was implemented as a Custom Function Unit (CFU) tightly coupled with a RISC-V processor within the CFU-Playground framework. This co-design preserves software flexibility while enabling dedicated acceleration of the full DSC block. A three-stage pipelined hardware design, supported by specialized memory buffers and on-the-fly padding logic, allows each output pixel to be computed completely in a single stream without intermediate storage.

FPGA validation on a low-cost Artix-7 board demonstrated up to a \textbf{59× speedup} over the baseline software implementation and a \textbf{20–30× improvement} over the original CFU-Playground accelerator. ASIC synthesis results at 40~nm and 28~nm confirmed the scalability and efficiency of the architecture, achieving compact area utilization and sub-watt power operation suitable for edge devices.

A detailed comparison with prior DSC accelerators showed that our fused architecture achieves the largest reduction in memory traffic—about \textbf{87\%}—while using fewer resources than most high-performance FPGA implementations. Unlike partial-fusion or hybrid-buffer approaches, our design performs full Ex$\rightarrow$Dw$\rightarrow$Pr fusion with zero intermediate buffering, bridging the gap between high-performance dataflow architectures and TinyML-level resource constraints.

In summary, this work demonstrates that careful hardware–software co-design, dataflow fusion, and fine-grained pipelining can deliver significant energy and performance benefits for embedded AI systems. Future work will focus on extending the fused dataflow concept to other neural network operations and integrating dynamic scheduling techniques to further improve utilization across diverse TinyML workloads.
%-----------------------------------------------------------
 \section*{Acknowledgment}

\bibliographystyle{IEEEtran}
\bibliography{references}

\end{document}